\documentclass[aps,pra,showpacs,twoside,twocolumn,10pt,superscriptaddress,nofootinbib,reprint,floatfix,%
amsmath,amssymb,%
floatfix,
]{revtex4-2}
\usepackage{hyperref}
\usepackage[T1]{fontenc} % needed for \DJ in bib entry
\usepackage{nicefrac}
\usepackage{graphicx}% Include figure files
\usepackage{dcolumn}% Align table columns on decimal point
\usepackage{bm}% bold math
\usepackage{xcolor}
\usepackage[ruled,vlined,linesnumbered]{algorithm2e}
\usepackage{setspace}
\usepackage[noend]{algpseudocode}
\usepackage{svg}

\begin{document}

\title{Using copies to improve precision  in continuous-time quantum computing}
\author{Jemma Bennett} 
\email{jemma.e.bennett@durham.ac.uk}
\affiliation{Physics Department, Durham University, South Road, Durham, DH1 3LE, UK}
\author{Adam Callison}
\affiliation{Department of Physics and Astronomy, UCL, Gower Street, London, WC1E 6BT, UK}
\author{Tom O'Leary} 
\affiliation{Physics Department, Durham University, South Road, Durham, DH1 3LE UK}
\affiliation{Department of Physics, Clarendon Laboratory, University of Oxford, Parks Rd, Oxford, OX1 3PU, UK}
\author{Mia West}
\affiliation{Physics Department, Durham University, South Road, Durham, DH1 3LE UK}
\author{Nicholas Chancellor}
\affiliation{Physics Department, Durham University, South Road, Durham, DH1 3LE UK}
\author{Viv Kendon}
\email{viv.kendon@strath.ac.uk}
\affiliation{Physics Department, Durham University, South Road, Durham, DH1 3LE UK}
\affiliation{Department of Physics, University of Strathclyde, Glasgow, G4 0NG, UK}
\date{\today}

\begin{abstract}

In the quantum optimisation setting, we build on a scheme introduced by Young et al [PRA 88, 062314, 2013], where physical qubits in multiple copies of a problem encoded into an Ising spin Hamiltonian are linked together to increase the logical system's robustness to error.
We introduce several innovations that improve this scheme significantly.
First, we note that only one copy needs to be correct by the end of the computation, since solution quality can be checked efficiently.
Second, we find that ferromagnetic links do not generally help in this ``one correct copy'' setting, but anti-ferromagnetic links do help on average, by suppressing the chance of the same error being present on all of the copies.
Third, we find that minimum-strength anti-ferromagnetic links perform best, by counteracting the spin-flips induced by the errors.
We have numerically tested our innovations on small instances of spin glasses from Callison et al [NJP 21, 123022, 2019], 
%and on small spin chains, 
and we find improved error tolerance for three or more copies in configurations that include frustration.
Interpreted as an effective precision increase, we obtain several extra bits of precision for three copies connected in a triangle.
This provides proof-of-concept of a method for scaling quantum annealing beyond the precision limits of hardware, a step towards fault tolerance in this setting.

\end{abstract}

\maketitle

%--------------------%
\section{Introduction}

In order to perform a quantum computation and benefit from the predicted quantum speedup \cite{Bernstein1993, Simon1994, Deutsch1992, Shor1994a}, it is necessary to maintain coherent quantum states for relatively long periods of time in the presence of noise. 
For quantum computers to be viable, ways of suppressing and correcting the errors caused by the noise are therefore necessary. 
For gate-based quantum computing, a framework for overcoming these obstacles was established with the development of fault-tolerant quantum error correction (QEC) in \cite{Shor1996, Knill1996} and its scalability in \cite{Knill1998, Aharonov2006}. Redundancy is created by entangling multiple qubits together. To detect errors, rather than directly measuring the data qubits, auxiliary qubits are introduced such that measuring the auxiliary qubits tells us about the parity between the data qubits.  From these measurements, the qubit (or qubits) on which an error has occurred may be identified and then corrected.
These error correction methods have also been adapted to measurement-based quantum computing \cite{Rauss2001} using techniques such as lattice surgery \cite{Horsman2012}.  Many refinements and improvements continue to be developed in the gate model setting (for a recent review, see \cite{Girvan2021}), in particular to reduce the number of physical qubits required per logical qubit.

For other types of quantum computing, error correction is less well-developed.  In this paper, we focus on continuous-time quantum computing, which includes adiabatic quantum computing, quantum annealing, and continuous-time quantum walks. 
Each of these start with the quantum system in an initial state that is the easily prepared ground state of a simple Hamiltonian $\hat{H}_0$.  The system is then driven over time into the ground state of a Hamiltonian $\hat{H}_P$ that encodes the problem to be solved. The Hamiltonian $\hat{H}(t)$ that carries out the computation can typically be written,
\begin{equation}
    \hat{H}(t) = A(t)\hat{H}_0 + B(t)\hat{H}_P,
\end{equation}
where $A(t)$ and $B(t)$ are the time-dependent control functions which differ for each type of continuous-time quantum computing. 

In adiabatic quantum computing (AQC), $A(t)$ is varied from 1 to 0 and $B(t)$ from 0 to 1, slowly and smoothly. As the system is evolved from $\hat{H}_0$ to $\hat{H}_P$, it remains in the ground state with high probability \cite{Farhi2000, Farhi2001}.
Quantum annealing also evolves the system from $\hat{H}_0$ to $\hat{H}_P$, however, instead of relying on the condition of adiabaticity, it uses other effects such as cooling towards the ground state \cite{Finnila1994, Kadowaki1998}. Terminology around quantum annealing is not universally agreed upon, but generally it is considered to encompass both cases where environmental dissipation plays a role in the computation, and where closed system effects dominate but evolution is faster than adiabatic (known as diabatic quantum annealing \cite{Crosson2020}).
Continuous-time quantum walks evolve the system under a time-independent Hamiltonian with the constant functions $A(t) = \gamma$ and $B(t) = 1$, where $\gamma$ is a hopping rate between states, followed by measurement at a suitable time $t_f$ \cite{Farhi1998, Childs2004}. 
In the case of quantum walks and quantum annealing, the probability of finding the correct ground state will usually be significantly less than unity, and repetitions are used to increase this probability. 
These three continuous-time techniques can be thought of as extremal points in a space of hybrid methods that can be interpolated between \cite{Morley2017}, to find methods suited to a particular problem and hardware.

While continuous-time quantum computing can be affected by bit-flip and phase-flip errors induced by environmental noise, similar to gate-based quantum computing, the equivalent of gate errors appear as errors in controlling the Hamiltonians. These include limited resolution in the hardware controls that hence cannot implement sufficiently precise values of the fields and couplers, and limitations in the dynamical controls that enact the functions $A(t)$ and $B(t)$ as the anneal is performed.  

Scalable fault-tolerant QEC for continuous-time quantum computing has yet to be established and is subject to several caveats \cite{Sarovar2013, Young2014, Marvian2014}, in particular, that two-local commuting Hamiltonians are not sufficient for constructing ground subspace encodings \cite{Marvian2014}.  Despite the challenges, a variety of schemes for error correction and suppression in continuous-time quantum computing have been developed, generally referred to as Hamiltonian error suppression \cite{Crosson2020}. 
Most can be grouped into categories: 
energy penalty Hamiltonians \cite{Jordan2006, Bookatz2015, Marvian2016, Marvian2017}; 
dynamical decoupling \cite{Lidar2008, Quiroz2012, Ganti2014};
subsystem codes \cite{Jiang2017, Marvian2017a, Marvian2019, Bacon2001, Lidar2019}; 
continuous-in-time techniques \cite{Sarovar2005, Sarovar2013, Atalaya2020};
via qubit ensembles \cite{Mohseni2019}; the Zeno effect \cite{Paz-Silva2012}; and QAC \cite{Young2013, Pudenz2014, Vinci2015, Pudenz2015, Vinci2016a, Matsuura2016, Mishra2016, Matsuura2017a, Vinci2018, Matsuura2019a, Li2020}. In addition to these techniques for explicit error suppression and/or correction, quantum annealing may be carried out in some circumstances without error correction, as long as sufficiently many repetitions are implemented \cite{Brooke1999, Lanting2014, Denchev2016, Johnson2011a, Boixo2016a}.

In this paper, we develop a scheme to protect the ground state of an Ising problem Hamiltonian from errors due to lack of precision in implementing its fields and couplings. 
Our scheme is built on quantum annealing correction (QAC), first introduced in \cite{Young2013, Vinci2015}.
We introduce several innovations to the scheme in \cite{Young2013, Vinci2015}.  In QAC, $C$ copies of an Ising Hamiltonian are connected together in chain or grid structures via strong ferromagnetic links, acting similarly to a repetition code as used in gate-based quantum computing or classical error correction.
Indeed, since Ising Hamiltonians, and the problems encoded in them, are classical (solutions are represented by computational basis states), only bit-flip errors need to be considered to protect the problem Hamiltonian.
Phase errors may become important in the dynamics when the driver Hamiltonian $\hat{H}_0$ is used, but are not relevant to the error models in the setting we consider in this work.

Our innovations are as follows.
First, we note that in the setting where we are solving hard classical optimization problems, we only need one of the copies to provide a correct answer.  This is because it is efficient to compare the quality of the candidate solutions, and select the best one.  This is the strategy employed in practical implementations using multiple runs, regardless of any error correction techniques used.  This fundamentally changes the criteria for evaluating the performance of the error correction strategies, and thus identifies different strategies that work well.
Second, we use weak anti-ferromagnetic links to connect the corresponding qubits in the copies, rather than strong ferromagnetic links.  As we will show, anti-ferromagnetic links can suppress the probability of an error affecting all of the copies, while ferromagnetic links do not.  
Third, we ensure the anti-ferromagnetic links induce frustration in the system by connecting three copies in a triangle configuration (each logical qubit is formed from three physical qubits connected this way).  This improves the suppression of errors.

We have numerically tested our innovations on small instances of three problems: Sherrington-Kirkpatrick spin glasses as studied in \cite{Callison2019}; maximum independent set; and Ising spin chains with random couplings.
However, spin chains have been found to exhibit counter-intuitive statistical effects \cite{Chancellor2020b}, so will not necessarily provide a good guide for how other types of problems behave.
Understanding the behavior for Ising spin chains is important in relation to minor embedding techniques used to instantiate highly connected problems into less connected hardware.  
%\jemma{How is this?}
Anti-ferromagnetic chains were studied in \cite{Pudenz2014} and found to be particularly challenging for QAC, due to the domination of domain wall errors in this type of problem. Minor embedding employs chain like structures and hence reduces the performance of QAC.
%\jemma{Need to chain-ge this!} 

We find that three copies connected in a triangle using weak anti-ferromagnetic links can maintain the correct ground state on one or more of the connected copies, even when the errors have changed the ground state for a single problem instance.
We use a simple error model to represent limited precision in applying the fields and couplings in the Ising Hamiltonian.  Interpreting our results as improved precision,
using these three connected copies can gain several bits of effective precision in the fields and couplings that represent the problem.  Since larger problems require more precise settings, our results provide proof-of-concept that linked copies can be used to increase the size of the problem that can be solved on hardware with limited precision.  This is similar in spirit to error correcting codes removing gate errors in digital quantum computing, allowing longer and larger computations to successfully complete, but at a cost of using more qubits.  Our results open the door to a route to scalability and the equivalent of fault tolerance in a continuous-time quantum computing setting.
Our results are numerical, limited by classical computing capabilities, and are thus proof-of-concept, not proof.  Nonetheless, they are important for introducing and validating the ideas as a prelude to further development.

Our approach is complementary to the work developing nested QAC \cite{Vinci2016a, Vinci2018, Matsuura2019, Matsuura2019a}
and penalty qubit QAC \cite{Pudenz2014, Vinci2015, Pudenz2015, Matsuura2016, Mishra2016, Matsuura2017a, Pearson2019, Matsuura2019} where an odd number of copies are linked ferromagnetically, and the repetition code is most often decoded by majority vote, although `coin tossing' or energy minimization may also be used \cite{Vinci2015}.  
The mean field analysis in \cite{Matsuura2016} shows that, for simple Ising or $p$-spin ferro- or anti-ferro problem Hamiltonians, copies connected in complete graph structures (all to all couplers) can prevent closing of the minimum energy gap and for nested quantum annealing in \cite{Vinci2018}, reduce the effective temperature, or, equivalently, increase the effective energy scale, both of which mitigate against thermal and other environment-induced errors.

The paper is organized as follows.  
In section \ref{sec:back} we provide background to quantum optimization in a transverse Ising model setting, and introduce several concepts relevant for our error models. 
In section \ref{sec:numeth} we summarize the numerical methods we used to test our models.
In section \ref{sec:links} we describe how we determined the best link strength to use. Then in section \ref{sec:precision} we show how this leads to a significant improvement in the precision. In section \ref{sec:QWs}, we test how the dynamics of quantum walks interact with on our error suppression strategy. We conclude in section \ref{sec:conc}.

%----------------------------------%
\section{Background}\label{sec:back}

%-------------------------------------------------------%
\subsection{Transverse Ising Hamiltonian Optimization}\label{ssec:TIM_opt}

Classical optimization problems can be efficiently encoded \cite{Choi2010} into an $n$-qubit Ising Hamiltonian $\hat{H}_P$ of the form
\begin{equation} \label{eq:Is_H}
    \hat{H}_P = \sum_{j=0}^{n-1}h_j\hat{Z}_j + \sum_{j\ne k=0}^{n-1}J_{jk}\hat{Z}_j\hat{Z}_k,
\end{equation}
where the fields with strengths $h_j$ act on the $j$th qubit and the couplings with strengths $J_{jk}$ act between the qubits $j$ and $k$. The $\hat{Z}_j$ operators are tensor products
\begin{equation}\label{eq:Ztensor}
    \hat{Z}_j = \left( \bigotimes^{j-1}_{r=0} \mathbb{I}_2 \right) \otimes \hat{Z} \otimes \left( \bigotimes^n_{j+1} \mathbb{I}_2 \right),
\end{equation}
which act non-trivially with a Pauli-Z operator $\hat{Z}$ on only the $j$th qubit.
The strengths of the fields $h_j$ and couplings $J_{jk}$ will be limited in range by the hardware capabilities.  To model this, we use a range restricted to the interval [-1,1] for both in our numerical simulations.

%---------------------------------------%
\subsection{Definitions of problems used}\label{ssec:spin_defs}

The three different problems we use in this work are:

\paragraph{A data set of Sherrington-Kirkpatrick (SK) spin glass instances} which were generated for research reported in \cite{Callison2019} and made available here \cite{Chancellor2019a}.  Finding the ground state of SK spin glasses is NP-hard, and also `uniformly hard', meaning almost all problem instances are hard at large sizes.  
Sherrington-Kirkpatrick spin glasses \cite{Sherrington1975} are defined by,
\begin{equation}
    H_{SK} = -\frac{1}{2} \sum_{(j \neq k)=0}^{n-1}J_{jk}S_jS_k,
\end{equation}
    where $S_j$ are the classical spins ($S_j \in \{ -1, 1 \}$) and the couplings $J_{jk}$ are drawn independently from a normal (Gaussian) distibution with mean zero and standard deviation $\sigma$. Single-body field terms, $\sum^{n-1}_{j=0}h_jS_j$ are added to break the spin inversion symmetry, where $h_j$ are the field strengths and are also drawn independently from the same normal distribution as the couplings. In order to map this to the quatum Ising model, the classical spin variables $S_j$ are simply mapped to Pauli-$Z$ operators and the problem Hamiltonian becomes,
    
    \begin{equation}
         \hat{H}_{SK} = -\frac{1}{2}\sum_{j\neq k=0}^{n-1}J_{jk}\hat{Z}_j\hat{Z}_k -
         \sum_{j=0}^{n-1}h_j\hat{Z}_j.
    \end{equation}

\paragraph{The maximum independent set} 
(MIS) is an NP-hard problem that is well-studied in computer science/graph theory. An independent set is a set of vertices in a graph $G$ that are not adjacent (i.e. not connected by an edge). The maximum independent set (MIS) is the maximum possible size of this set for a given graph $G$. A 5 vertex problem is used as an example to show the effect of our error suppression method in section \ref{ssec:err_sup}.

\paragraph{Ising spin chains} have qubits linked with couplings $J_{jk}\ne 0$ when $k=j+1$. They are similar to the multi-qubit variable mappings used on D-Wave machines for minor embedding 
%%%
\cite{choi10a, choi08a,minorminer}. 
%%%
However, unlike the SK spin glasses, finding the ground state of spin chains can be solved efficiently classically, and they have been shown to exhibit counter-intuitive statistical effects \cite{Chancellor2020b}, so they are not suitable for gaining insight into the behavior of more general types of problem Hamiltonians.  Nonetheless, they are important for real hardware architectures (e.g., minor embedding), and their simplicity is useful in illustrative diagrams.

%--------------------------------------------------------------%
\subsection{Error models and precision}\label{ssec:prec}

In any real hardware, there is a limit to how precisely a field $h_j$ or coupling $J_{jk}$ can be set.  The theoretical model in equation \eqref{eq:Is_H} allows these to be real numbers, but in practice we only have a fixed number of possible values available.  Thus, we cannot in general represent the problem Hamiltonian exactly, the limited resolution of the hardware will be a potential source of error, even before we carry out the computation.  This issue was first recognised in the AQC setting by \citeauthor{Young2013} \cite{Young2013}.
There are many other sources of errors in real hardware, but in this work, we focus just on the effects of limited precision.  This is a simple and easy to understand model that is of practical importance.
A deeper understanding of one type of error is useful for designing error mitigation strategies that can handle realistic situations with multiple sources of error.
\begin{figure}[!tb]
    \centering
    \includegraphics[width=0.4\textwidth]{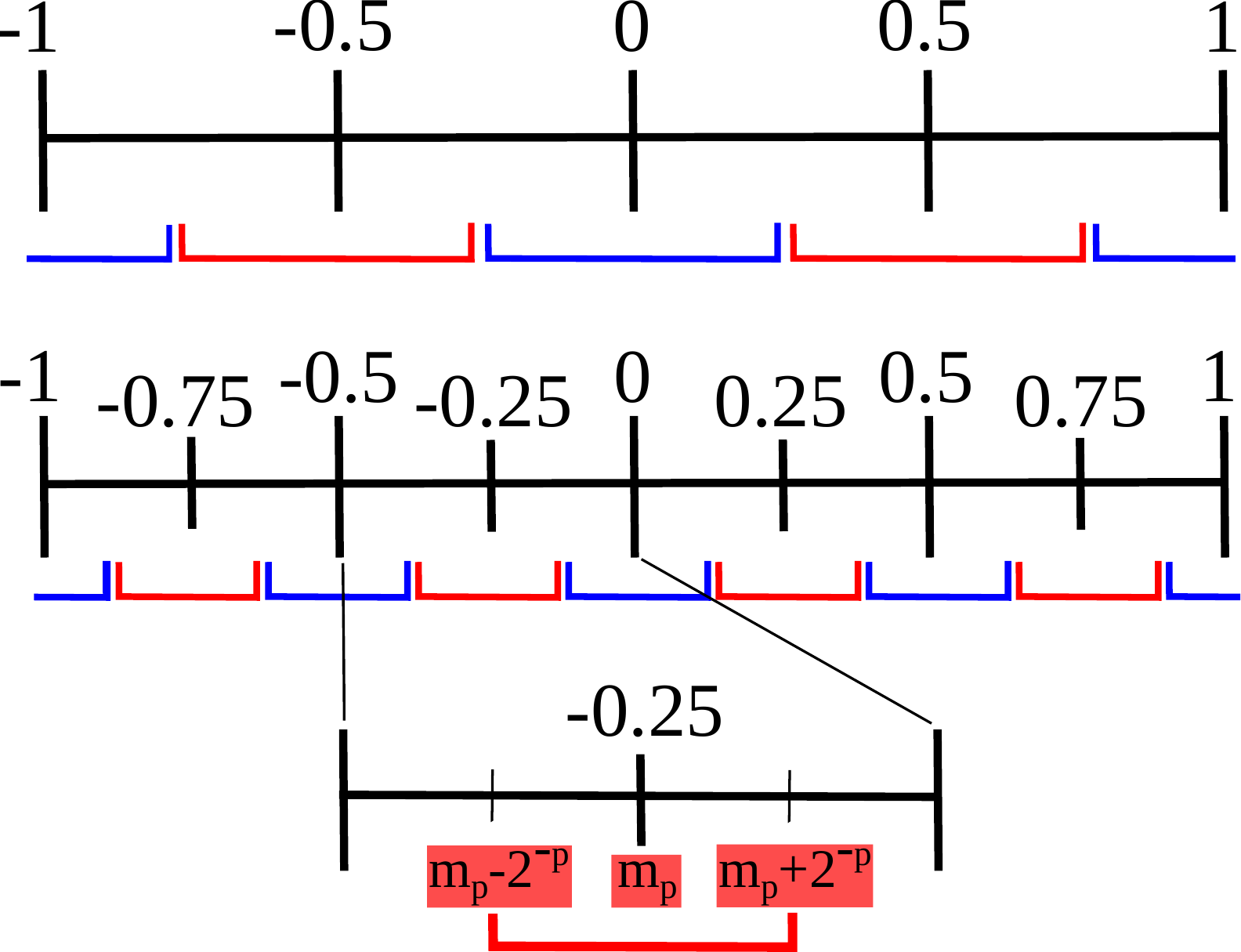}
    
    \caption{Two number lines showing the resolution available between -1 and 1, from $p=2$ bits ($2^2=4$ values) and $p=3$ bits ($2^3=8$ values). The uncertainty of each of the values is shown below the line in alternating red and blue delineations of the divisions. Underneath the number line  for 3 bits, the area between 0 and -0.5 is expanded to show the location of the midpoint $m_p$ and the upper and lower limits: $m_p \pm 2^{-p}$. }
    \label{fig:n_lines}
\end{figure}

Our model of limited resolution, quantified by the precision $p$, divides the range $[-1,1]$ into $2^p+1$ possible values (the resolution).  The natural way to do this for periodic boundary conditions is as shown in figure \ref{fig:n_lines}.  The divisions shown below the line are labeled by their midpoints $m_p$, and zero is one of the midpoints. The endpoints of the division labeled zero are $\pm2^{-p}$, with $p=2$ for the upper line and $p=3$ for the lower line.  The precision $p$ is thus the number of classical bits needed to represent the possible values in the range $[-1,1]$.
Of course, we do not have periodic boundary conditions for the fields and coupling strengths $h_j$ and $J_{jk}$.  The natural way to divide a range with endpoints is to shift the divisions in figure \ref{fig:n_lines} so that the midpoints are $\pm2^{-p}$, etc.  However, this means there are two divisions that could represent approximately zero, and for reasons which will become clear when we present our results, we prefer to have only have one division representing approximately zero.  Hence, we use the model in figure \ref{fig:n_lines}, but with two ``half divisions'' at each end.

Operationally, limited precision means that when we set a parameter to $m_p$, the midpoint of the division, the actual value obtained in the hardware could be anywhere between $m_p\pm2^{-p}$, with some probability distribution.
More generally, if we set a value $x$ such that  $m_p-2^{-p}\le x\le m_p+2^{-p}$, we instead obtain $y$ such that $m_p-2^{-p}\le y\le m_p+2^{-p}$ with probability $P(y|x)$.
If $P(y|x)$ is the uniform distribution on the interval $m_p\pm2^{-p}$, the average error in $x$ scales quadratically in $x$, as $\varepsilon_u = 2^{p-1}\{(x-m_p)^2+2^{-2p}\}$, ranging from $\nicefrac{1}{2}\,2^{-p}$ for $x=m_p$ to $2^{-p}$ for $x$ at the endpoints of the division, with the average over $x$ being $\langle\varepsilon_u\rangle_x = \nicefrac{2}{3}\, 2^{-p}$.

Instead of averaging over the error distribution $P(y|x)$, for numerical simulation purposes it is useful to use a deterministic error model.  This simplifies our numerical simulations: instead of averaging over many randomly chosen $y$ for each $x$, we use a single randomly chosen $y$ for each $x$ at precision $p$.  This also means that using repetitions will not increase the probability of a successful computation, if the randomly chosen $y$ gives the wrong ground state.  This allows us to separate the effects of repeat runs from the effects of our error mitigation strategies.

Another natural model is to round to the midpoint $m_p$ in each division, a model of a dial which only has discrete settings available.
The error in the mid-point model scales linearly, $\varepsilon_m = |x-m_p|$, ranging from zero at the midpoint to $2^{-p}$ at the endpoints, with the average over $x$ being $\langle\varepsilon_m\rangle_x = \nicefrac{1}{2}\, 2^{-p}$.  This underestimates the error compared to the uniform error model, especially when the exact $x$ is close to $m_p$.  
However, especially when the precision $p$ is low, using the mid-point model means there is a high likelihood of terms within the Hamiltonian cancelling, leading to a high level of degeneracy in the ground state of the Ising model instance with error.
This high level of degeneracy is not realistic for experimental hardware, for example in flux qubit quantum annealers, as programmable coupling values are not evenly spaced \cite{VanDerPloeg2007}.  The high level of degeneracy also causes problems in numerical simulation, if the degenerate ground states include the correct solution.  Hence, we use the random error models for most of our simulations.
Real hardware is likely to have a non-uniform distribution for $P(y|x)$, so neither the random nor the mid-point model is preferred on realistic grounds.

\begin{figure}[!tb]
    \centering
    \includegraphics[width=1.0\columnwidth]{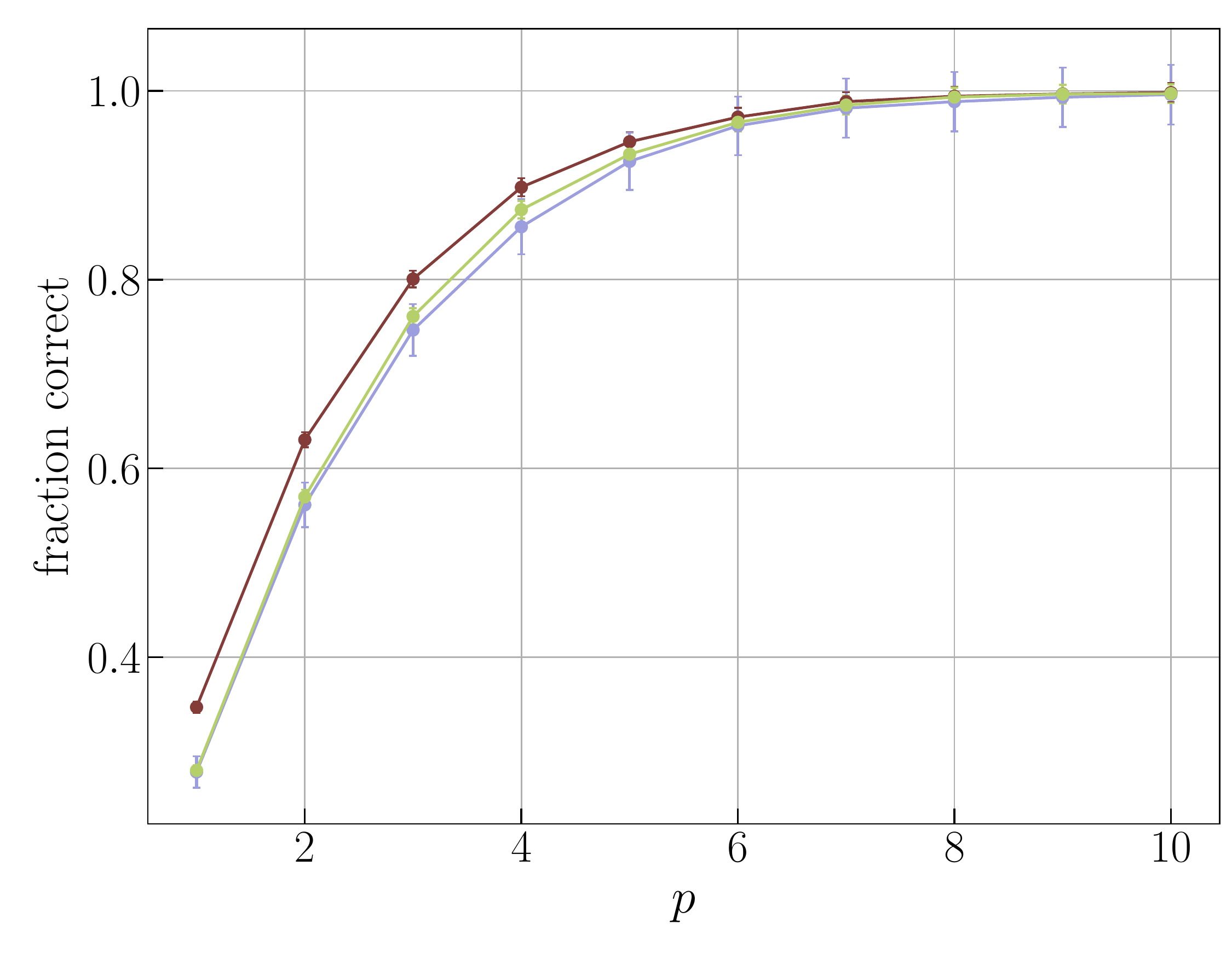}
    \caption{ Fraction of instances with the correct ground state vs precision $p$ for $10^4$ instances of 5 qubit SK spin glasses, rounding each of the parameters to the midpoints (red), a single sample of random uniform errors on each of the parameters (green) and $10^3$ instances, averaging over 10 samples of uniform random errors on each of the parameters (blue).}
    \label{fig:rand_vs_prec}
\end{figure}

We illustrate the difference between the error models in figure \ref{fig:rand_vs_prec}, by calculating the fraction of instances with correct ground states when rounded to a given precision $p$.
The data plotted have been computed using the 5 qubit instances from \cite{Chancellor2019a}. Each of the $10^4$ instances was subjected to one of the error models, which generated an approximated Hamiltonian. The ground state of this approximated Hamiltonian was then compared to the ground state of the exact Hamiltonian. Each of the $10^4$ instances has been processed with three different error models.  First for the midpoint model, where each field $h_j$ and coupling $J_{jk}$ is rounded to the nearest midpoint value allowed by the precision $1\le p\le 10$. 
Second for the deterministic random model, where a single new instance was generated by drawing new values for each $h_j$ and $J_{jk}$ uniformly at random, across the division around the nearest midpoint value, for each precision $1\le p \le 10$. Third for the uniform random model, where instead of just one, 10 new instances were generated in the same way as for the deterministic random model.
The fraction retaining the correct ground state is obtained by averaging over all the data for each error model and precision.
As predicted, the midpoint model (red) finds the correct ground state slightly more often than the uniform random model (blue) and the deterministic random model (green) at low values of $p$. The differences between the mid-point, deterministic random and uniform random method are small, and the qualitative behavior is the same for all three models.

Figure \ref{fig:rand_vs_prec} also shows the typical size of the errors introduced by reducing the precision.  For the small sizes we study (up to 9 qubits, limited by computational capabilities), to break more than 5\% of the instances requires reducing the precision to around $p\lesssim n$ where $n$ is the number of qubits.  This makes sense intuitively, because $n$ qubits can only represent $2^n$ different outcomes, meaning that the average gap between adjacent energy eigenstates will be of order $2^{-n}$, and errors smaller than this are unlikely to change the ground state.

%-----------------------------------%
\subsection{Error suppression method}\label{ssec:err_sup}

As we have shown in figure \ref{fig:rand_vs_prec},
for precision $p \lesssim n$, the ground state changes in a significant fraction of the spin glass instances, meaning that a computation correctly finding the ground state will not always solve the original problem.
Our aim is to find a way to reduce these errors by using several copies, like a repetition code. 
In the uniform random error model, repeating the computation has a chance of finding the correct ground state in some of the attempts, as sometimes the incorrect settings will be close enough to the true values.  However, in the midpoint and deterministic random models the error is deterministic, standalone repeats will not help, unless there are degenerate ground states that include the correct solution.  Using the midpoint or deterministic random model thus provides a tougher test for our error correction methods: none of the improvement is due to repetition alone.  

Instead of repeat runs, we use several copies at the same time, and link the corresponding qubits to each other with coupling strength $J_F$.
The Hamiltonian at precision $p$ for the system of $C$ Ising model copies with each set of corresponding qubits connected according to a graph $G$ can be written as
\begin{align} \label{eq:Is_H_cps}
    \hat{H}_p^{(C)} =& \sum_{c=0}^{C-1}\left\{\sum_{j=0}^{n-1}   h_j^{(p)} \hat{Z}_{j}^{(c)}
      + \sum_{j\neq k=0}^{n-1}  J_{jk}^{(p)}\hat{Z}_{j}^{(c)}\hat{Z}_{k}^{(c)}\right\} \nonumber\\
      -& \sum_{j=0}^{n-1} \sum_{c,c'\in G} J_F^{(p)} \hat{Z}_{j}^{(c)}\hat{Z}_{j}^{(c')},
\end{align}
where $c,c'\in G$ means copies $c$ and $c'$ correspond to endpoints of an edge in graph $G$.

When $J_F^{(p)} >0$ (ferromagnetic links, as used in \cite{Young2013, Pudenz2014, Vinci2015, Pudenz2015, Matsuura2016, Mishra2016, Vinci2016a, Matsuura2017a, Vinci2018, Pearson2019, Matsuura2019a}),
the final term in this Hamiltonian makes it energetically favourable for the qubits in each of the copies to align.
For the exact Hamiltonian, all of the copies in the system will thus have the correct ground state, i.e. the links do not introduce any extra errors. 
Ferromagnetic links provide an energy barrier which suppresses bit-flip errors. However, if a bit-flip error does occur in one or more copies, the ferromagnetic links tend to propagate this error to all the other copies, and to neighboring qubits within copies, potentially making them all incorrect.
For random errors, if the error propagation is not too extensive, majority vote decoding can potentially remove the effects of the error
\cite{Pudenz2014, Vinci2015, Pudenz2015, Matsuura2016, Mishra2016, Vinci2016a, Matsuura2017a, Vinci2018, Pearson2019, Matsuura2019a}.
However, for the more deterministic sort of errors introduced by lack of precision, this strategy is less effective.

\begin{figure}[!tb]
    \centering
    \includegraphics[width=0.315\columnwidth]{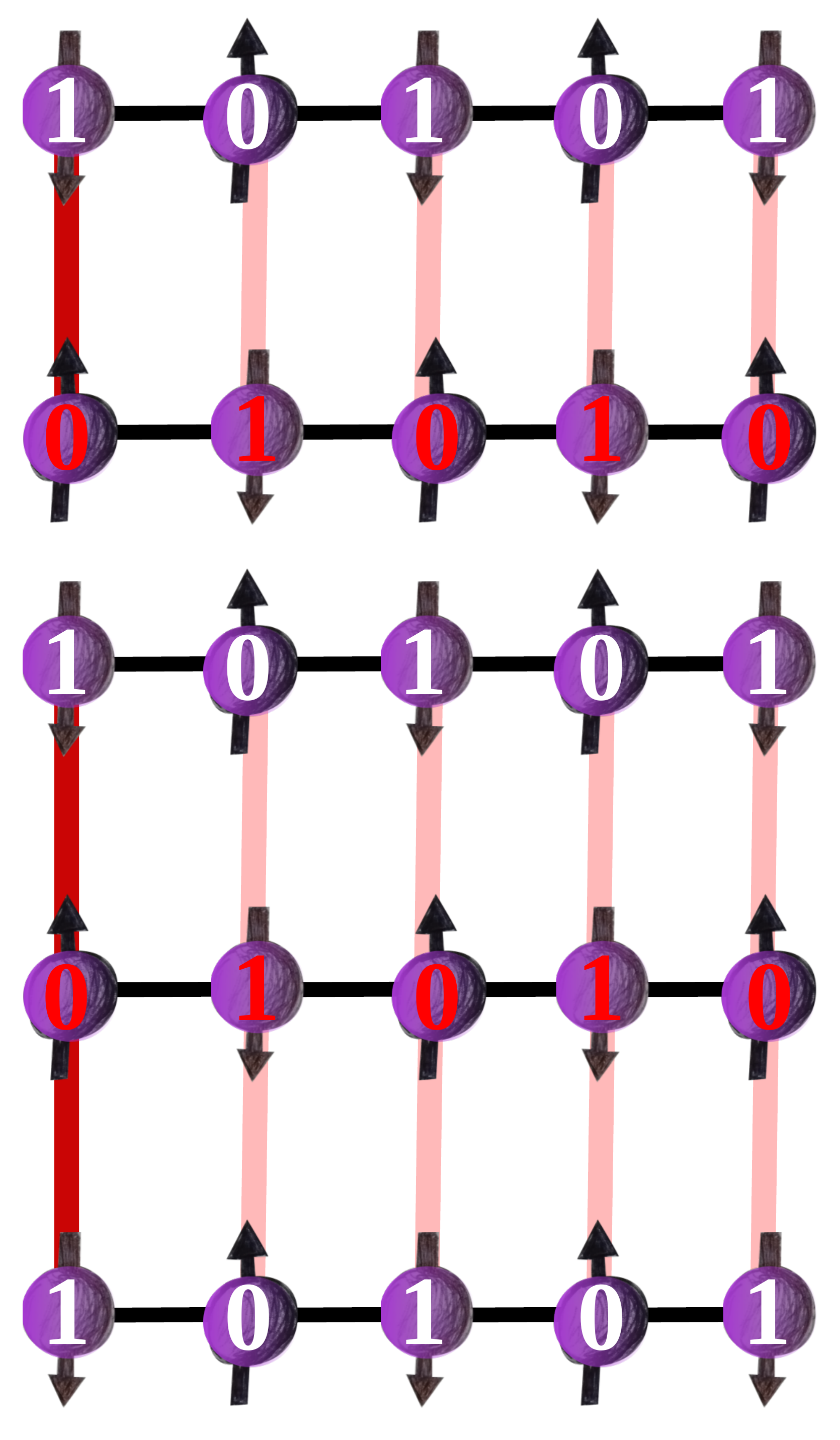}
    \includegraphics[width=0.4\columnwidth]{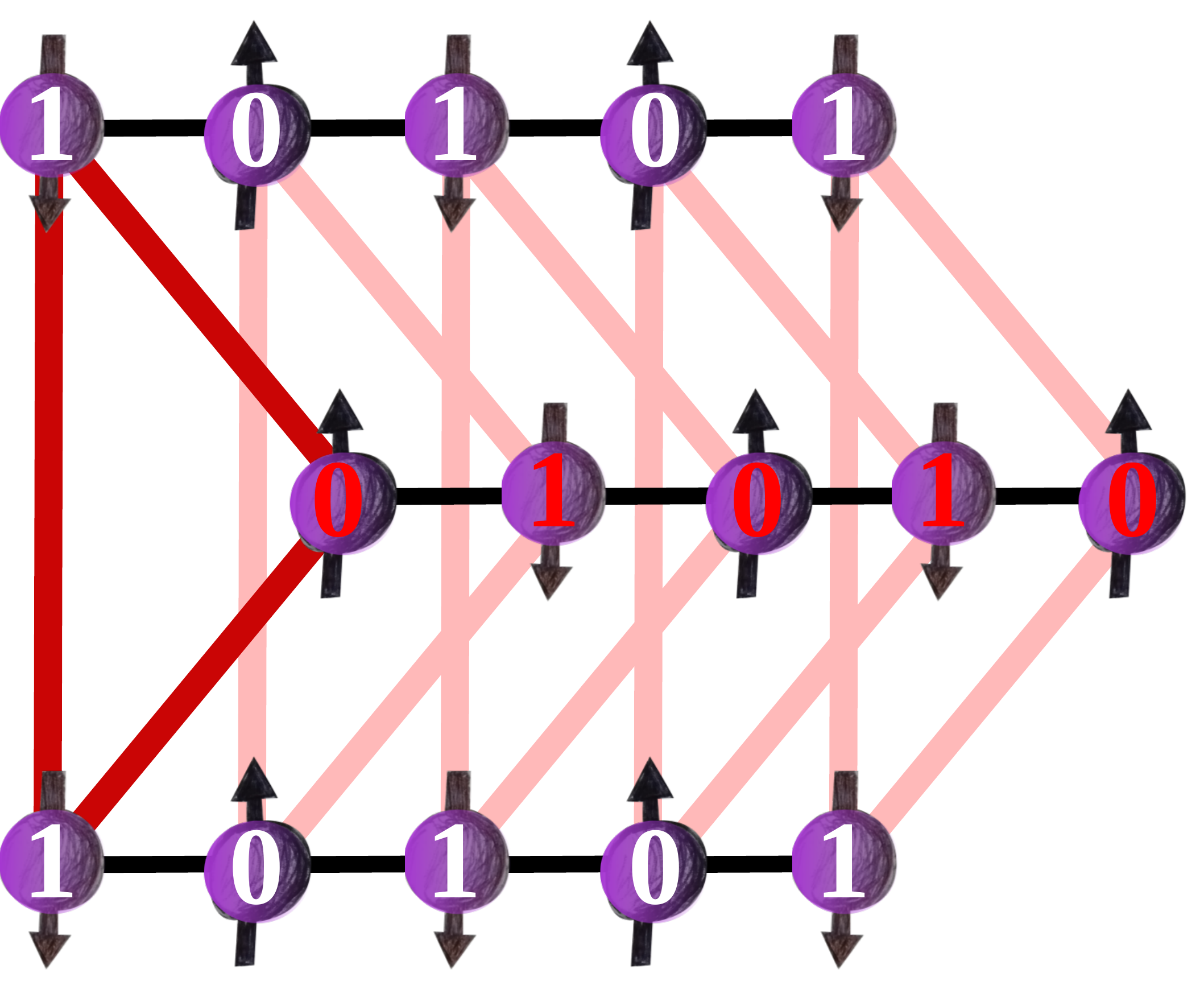}
    \caption{Diagram of two and three copies of a 5 qubit Ising chain (thin black links) with anti-ferromagnetic links (red/pink links) in a chain (left) configuration and three copies of the same Ising chain with anti-ferromagnetic links in a triangle (right) configuration.   The red links identify the corresponding qubits on the left end of the chain; the pattern continues with the pink links for the other four qubits.}
    \label{fig:3Isings}
\end{figure}
On the other hand, when $J_F^{(p)} < 0$ (anti-ferromagnetic links), the final term in the Hamiltonian makes it energetically favourable for the qubits in the copies to be anti-aligned. This means that, even for the exact Hamiltonian, one or more of the copies is likely to have an incorrect ground state.  However, the opposing spins create an energy barrier which suppresses further bit-flip errors. 
Hence, if a bit-flip error occurs in a single copy, the anti-ferromagnetic links in the linked copies can prevent the error occurring on all copies, making it more likely a correct copy persists despite the bit-flip errors.
This is illustrated in figure \ref{fig:3Isings}, using spin chains for clarity in the diagrams.

The work presented in this paper investigates in detail how and when it is effective to use anti-ferromagnetic links between copies to reduce bit-flip errors, and estimates the level of error reduction achieved.
Due to computational limitations, we focus on $C=2$ and $C=3$.
Two copies have just one $J_F^{(p)}$ link connecting each pair of corresponding qubits. 
Three copies may be connected linearly in a chain, or in a triangle configuration, see figure \ref{fig:3Isings}. 
For copies connected in a linear chain configuration, the corresponding qubits in each copy can alternate to satisfy the anti-ferromagnetic links.  However, when the copies are connected in a triangle configuration, one of the anti-ferromagnetic links must connect copies which do not have opposing spins. This introduces frustration into the system, which can prevent a (bit-flip) error appearing in every copy.

To illustrate the effect of the anti-ferromagnetic links and frustration compared to the original Ising problem, table \ref{MIS_table} shows a $5$ qubit maximum independent set (MIS) problem with precision reduced to $p=1$ through $4$. The graph is represented by the black connections between the qubits and a qubit in state one (zero) corresponds to a vertex which is (is not) in the independent set.
The problem Hamiltonian is
\[
\hat{H}_{\text{MIS}} =-\sum_j(d_j-\frac{1}{2})\hat{Z}_j+\sum_{j,k\in \text{edges}} \hat{Z}_j\hat{Z}_k 
\]
where $d_j$ is the degree of (number of edges connected to) node $j$ and each edge is only counted a single time in the second sum.
\begin{table}[!bt]
\begin{tabular}{|l|l|l|l|l|}
\hline
        & $p\ge4$ & & $p=2$&\\
 $p \quad$ & \includegraphics[width=0.15\textwidth]{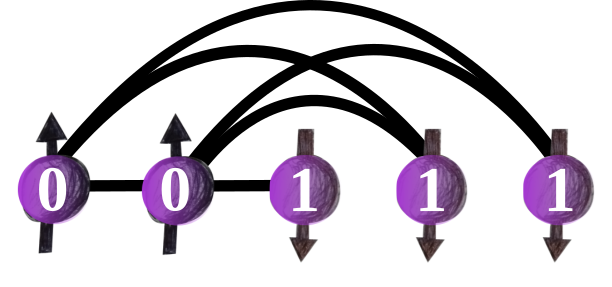} &
 \# \includegraphics[width=0.02\textwidth]{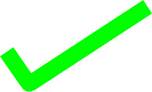}          &
 \includegraphics[width=0.18\textwidth]{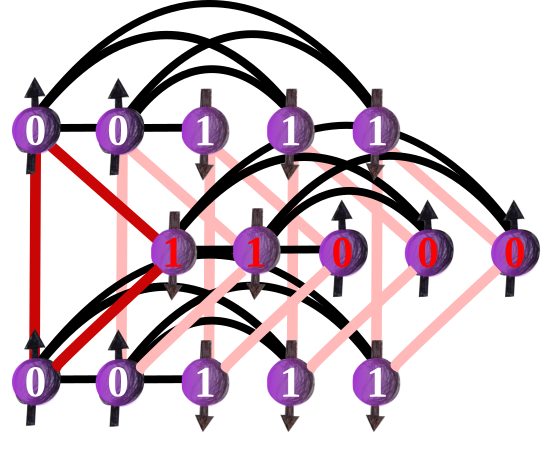}             &
 \# \includegraphics[width=0.02\textwidth]{Figures2/MIS/green_tick.png}\\ \hline
1       &   00\textcolor{red}{000}                    & 0  &  00\textcolor{red}{0}1\textcolor{red}{0}, \textcolor{red}{11}1\textcolor{red}{0}1, 00\textcolor{red}{0}1\textcolor{red}{0}               & 0  \\ \hline
2       &   \textcolor{red}{1}01\textcolor{red}{00}   & 0  &    
00111, \textcolor{red}{11000}, 00111                  & 2  \\ \hline
3       &   00\textcolor{red}{000}                    & 0  & 
00111, \textcolor{red}{1}0\textcolor{red}{000}, 00111 & 2  \\ \hline
4       &   00111                                     & 1  &  
00111, \textcolor{red}{1}0\textcolor{red}{000}, 00111 & 2  \\ \hline
\end{tabular}%
%}
\caption{Table comparing the numbers of correct ground states, for a single copy of a 5 qubit MIS Ising problem versus three connected copies, at precisions 1, 2, 3, and 4. For each ground state, correct qubits are shown in black and incorrect qubits in red.}
\label{MIS_table}
\end{table}
The single copy behavior for different precisions is shown in the left hand column.  The ground state is incorrect for $p<4$, with the incorrect qubits shown in red below the diagram.
The right hand column shows the behavior of three copies connected with $-J_F^{\text{min}}$ anti-ferromagnetic links, where $J_F^{\text{min}} $ is the smallest possible non-zero value of $J_F$ allowed at that precision.
The error model used is the midpoint model (deterministic errors).
Except for $p=1$, at each 
$p$, two out of the three copies have the correct ground state, despite the single copy being incorrect for $p=2$ and $3$.  This shows that our error suppression method works for this example.  Even with some incorrect copies, it is easy to check which of the three bit strings provides the best candidate solution to the MIS problem, by calculating (classically) the energy with the exact Hamiltonian.

%-------------------------------------------%
\section{Numerical methods}\label{sec:numeth}

Before presenting our main results using the \cite{Chancellor2019a} spin glass data set, we outline our numerical methods.  We used numerical simulation to obtain most of our results, run on desktop workstations and on HPC facilities based at Durham University, and at Oxford University via the UK Quantum Technology Quantum Computing and Simulation Hub. 
Our code was written in Python3 \cite{Python} from the Anaconda platform \cite{Anaconda}, with much of the editing and initial runs done using the Pycharm IDE \cite{Pycharm}. NumPy \cite{Numpy} and pandas \cite{Pandas} packages were used for data processing. Plotting was done using matplotlib \cite{matplotlib}.
Where relevant, figures have error bars, although the size of these is often too small to be seen.  For data sets of $10^4$ instances, the expected error in the averages is around 1\%.

We used sizes $5 \leq n \leq 9$ SK Ising spin glass data sets of $10^4$ instances per size, from data archive: \cite{Chancellor2019a}.
These have couplings $J_{jk}$ and fields $h_j$ drawn from a normal distribution with standard deviation $\omega_{SK}$, an arbitrary energy unit set to $\omega_{SK}=1$. 
%%%
For our error models, we require $J_{jk}$ and fields $h_j$ to be strictly within the range $[-1, 1]$.  Hence, 
we re-scaled so that the maximum and minimum values align with $\pm1$, which is the optimal mapping of the problem to a limited range of possible settings. 
Operationally, an equivalent rescaling needs to be done to fit problems into the range of values available in real hardware.

The ground states of these problems were found using a classical branch and bound algorithm. 
We quantify the effectiveness of our error suppression by comparing the fraction of problems successfully solved across the whole data set for a given size and precision.
We define \emph{fraction correct} as the number of instances which still have at least one correct ground state (out of the several linked copies), divided by the total number of instances used from the data set.

In section \ref{sec:QWs}, the effect of quantum walk dynamics on our error suppression method was analysed, using a quantum walk simulation code developed for use in \cite{Callison2019, Callison2020}.  Following the definition in \cite{Callison2019}, we use a time average success probability, defined as,
\begin{equation}\label{eq:Time_av}
    \bar{P}(t, \Delta t) \equiv \frac{1}{\Delta t}\int^{t+\Delta t}_t dt_f P(t_f),
\end{equation}
where $t$ and $\Delta t$ are chosen to be large enough to ensure reasonable convergence towards the infinite time average. The four 5-qubit spin glass examples a), b), c) and d) used in figure \ref{fig:QW_ex} were also taken from data archive: \cite{Chancellor2019a}, and have the unique I.D.'s 
    `acyenjvndejjyhbcfmkjefgzjtjqjt',
    `aakxejqunlcpqhmnftnrckailrczyp',
    `aclwrzmpznazfkjktzcswfdxjprfth',
    `acjdimxkqejkngndlykgxntdtxrgij', respectively.

%--------------------------------------------%
\section{Link strength tests}\label{sec:links}

%----------------------------------%
\subsection{Optimal link strength}

\begin{figure}[!tb]
    \centering
    \includegraphics[width=0.5\textwidth]{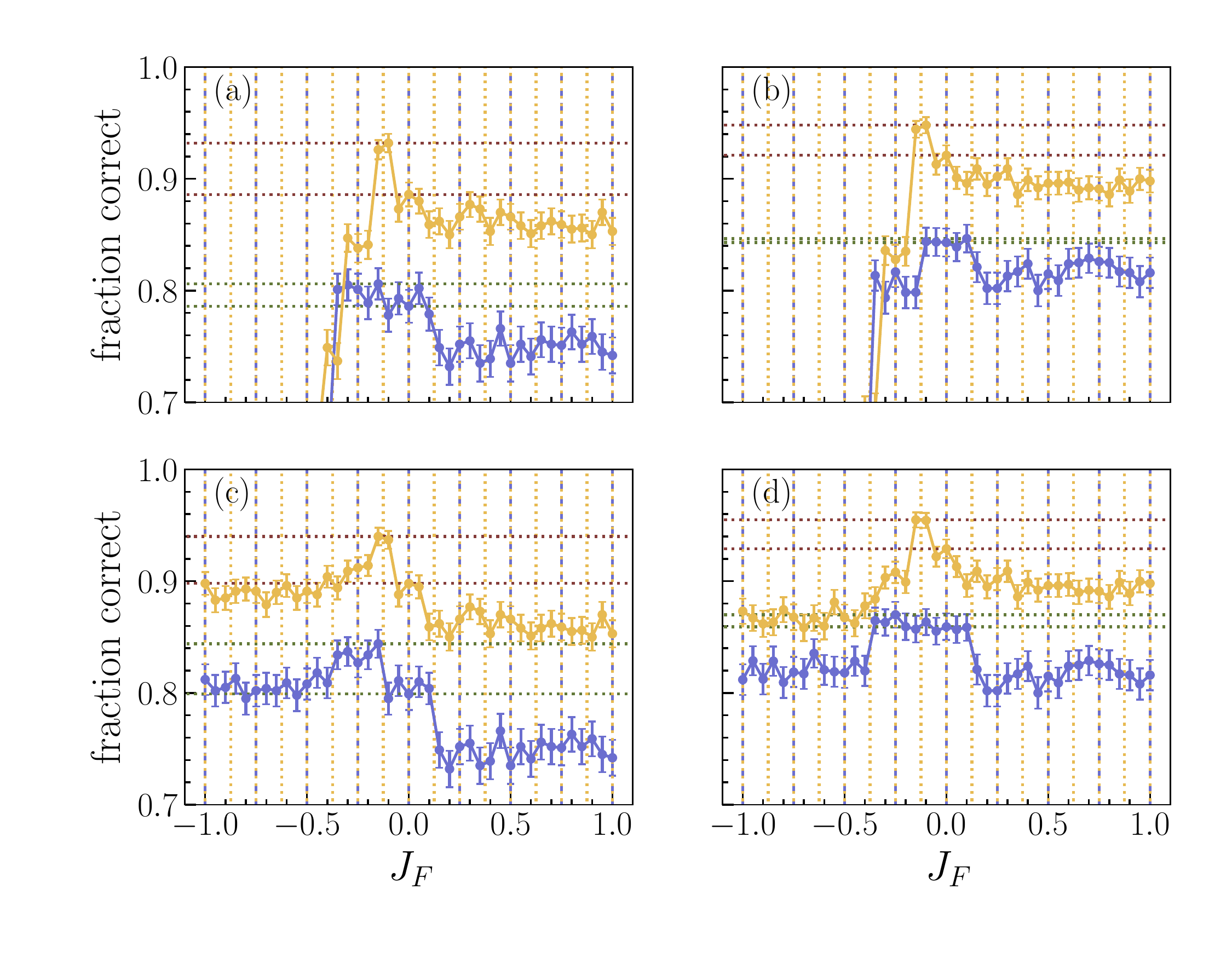}
    \caption{ Fraction correct against $J_F$ for $p=3$ (blue) and $p=4$ (yellow) for $10^3$ instances of 5 qubit problems for 2 copies (top row) and 3 copies in a triangle (bottom row) of spin glasses (left column) and spin chains (right column).
    Vertical dotted lines indicate the mid-points of each of the allowed divisions at that precision for $p=3$ (blue) and $p=4$ (yellow).  Horizontal dotted lines indicate the fraction correct for $J_F^{(p)}=0$ and the maximum fraction correct for any $J_F^{(p)}$, in green for $p=3$ and red for $p=4$.
    }
    \label{fig:4_panel_Jglass}
\end{figure}

Prior work \cite{Young2013,Pudenz2014, Pudenz2015,Vinci2015} using ferromagnetic links between qubit copies found that strong links are required, stronger than the typical field and coupling strengths in the problem Hamiltonian.  This is not favorable for scaling to larger sizes or more copies.  The scaling requirements were improved in subsequent work \cite{Matsuura2016, Vinci2016a, Matsuura2017a, Vinci2016a, Vinci2018, Matsuura2019, Matsuura2019a}, but the results still rely on optimising the ferromagnetic link strength, and are sub-optimal for problem Hamiltonians that use the full dynamic range of the hardware \cite{Vinci2018}.
In contrast, we find for anti-ferromagnetic links that the best improvement is obtained when $J_F$ is equal to $J_F^{(p)} = -2^{-p+1}$ in the mid-point error model and close to this value in the deterministic random error model. This is illustrated in figure \ref{fig:4_panel_Jglass} for the deterministic random model, averaged over $10^3$ instances of SK spin glasses (left) and spin chains (right). For consistency, we refer to the value where $J_F$ gains this best improvement as $J_F^{\text{min}}$ in both the mid-point and deterministic random model. 
Note that we tested intermediate values of $J_F^{(p)}$, to check that the value of $J_F^{\text{min}}$ is robust to variation within a division.
The choice of divisions in figure \ref{fig:rand_vs_prec} with only one representing approximately zero is now justified.  The other possible choice, with two divisions representing approximately zero, one with ferromagnetic character, the other with anti-ferromagnetic character, would have made it more difficult to extract the optimal value of $J_F^{\text{min}}$.

The four graphs show the fraction correct against link strength $J_F^{(p)}$ for two copies (top row) and three copies in a triangle configuration (bottom row).  Each graph shows results from both $p=3$ (blue) and $p=4$ (yellow), with dotted vertical lines indicating the mid-point values for each of the allowed divisions according to the precision.  
The results are similar for both spin chains and SK spin glasses, showing the robustness of the effect.  
Both spin glass and spin chain cases show an improvement for $J_F = J_F^{(\text{min})}$, for $p=4$ and all apart from the 2 copies of spin chains case show an improvement for $p=3$.  Horizontal dotted lines indicate the fraction correct for $J_F^{(p)}=0$, i.e., the baseline for any improvement, and the maximum fraction correct over all $J_F^{(p)}$ values, in green for $p=3$ and red for $p=4$.

For these relatively small 5 qubit problems, for two copies, for both spin glasses and spin chains, and for three copies in spin chains the improvement is only seen for $J_F^{(p)}\simeq J^{\text{min}}_F$ but for three copies of spin glasses in a triangle, some improvement persists for slightly larger (in magnitude) anti-ferromagnetic values of $J_F^{(p)}$.  These results show that there is more than one mechanism contributing to the observed improvements in the fraction correct. 

The improvement seen only for $J_F\simeq J^{\text{min}}_F$ is likely due to the extra link counteracting the impact of the imprecise field or coupling in the problem Hamiltonian.  In this case, the average error (see section \ref{ssec:prec}) is between $\nicefrac{1}{2}\,2^{-p}$ and $\nicefrac{2}{3}\,2^{-p}$, smaller but comparable with the strength of the minimum $J_F^{(p)}$.  Larger values of $J_F^{(p)}$ are thus likely to make things worse rather than better, by introducing more error than they counteract.  The steep fall in fraction correct for larger $J_F^{(p)}$ for two copies can be explained by this.
By a conceptually-similar mechanism to the one we show here, adding a small amount of thermal fluctuations has similarly been shown to mitigate against errors \cite{Nishimura2016}.

The improvements seen for three copies include the effects of frustration that tend to keep at least one copy correct, even when there are errors that should change the ground state.
The more gentle fall in fraction correct for three copies suggests this is significant, even when the link strength is not optimal.

\subsection{The deterministic error model}

\begin{figure}[!tb]
    \centering
    \includegraphics[width=0.8\columnwidth]{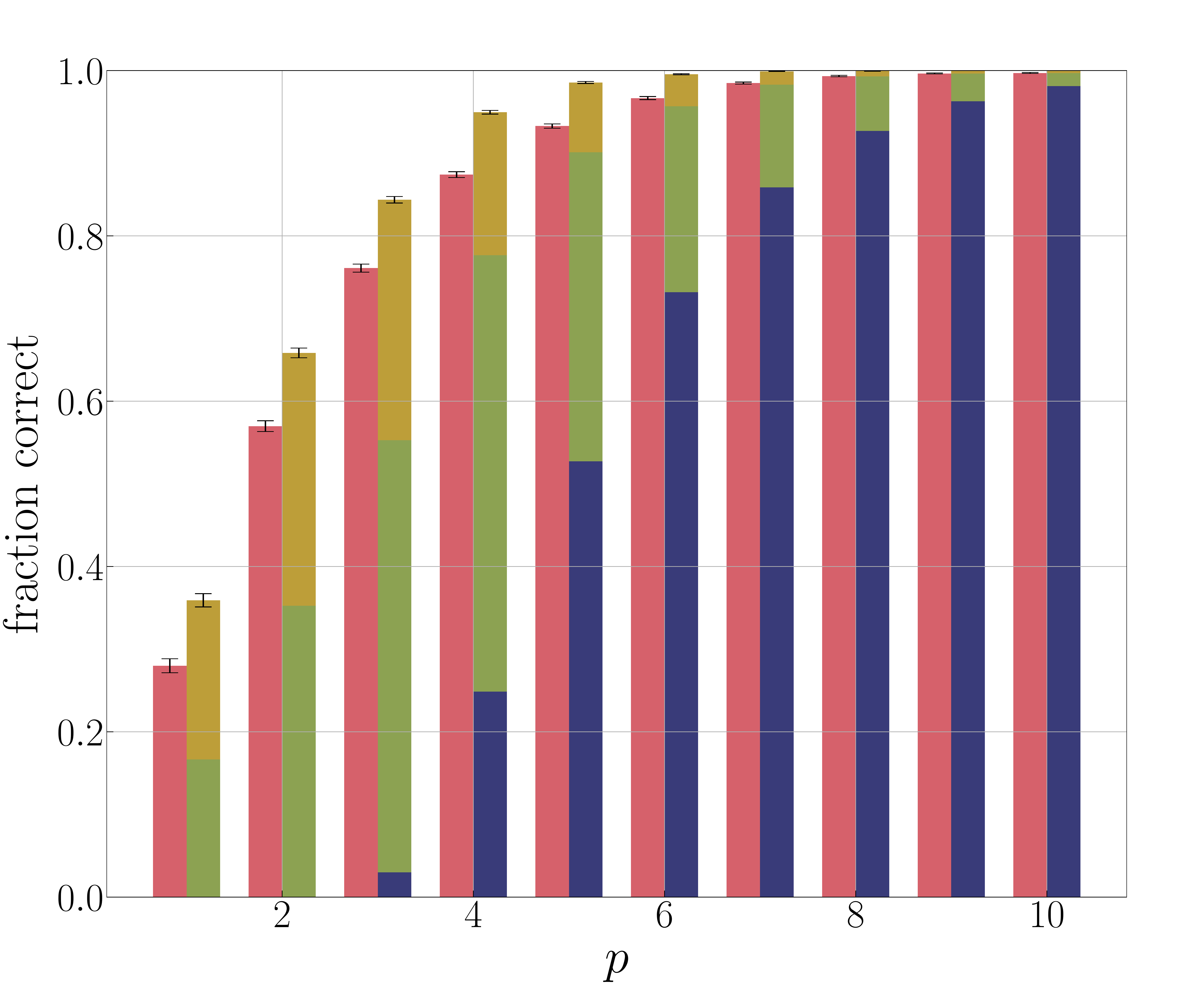}
    \caption{Fraction correct versus precision $p$ for $10^4$ instances of 5 qubit SK spin glasses. Left bars show fraction correct for single copies (red). Right bars  show fraction correct for 3 copies connected with links $J_F^{(p)}=J_F^{(\text{min})}$ split into all 3 correct (dark blue), 2 correct (green), 1 correct (yellow).}
    \label{fig:frac_corr_hist}
\end{figure}

 Having shown that there is a significant improvement, albeit on average rather than for every instance, by using anti-ferromagnetic links of strength $J_F^{(p)}\simeq J^{\text{min}}_F$ (the minimum allowed value by the precision), we now look in more detail at how the improvement is achieved.
 %consider how we can use this to improve the effective precision.  
We choose the case that provides the best improvement in the SK spin glasses, i.e., $J_F^{\text{min}}$, and three copies connected in a triangle to provide frustration. In figure \ref{fig:frac_corr_hist}, we compare the fraction correct versus precision $p$, averaged over $10^4$ instances of 5-qubit SK spin glasses, for a single copy and three connected copies. 
The light red bars (left) show the fraction correct averaged over all $10^4$ instances for 1 repeat of single copies for each precision $1\le p\le 10$. As there is little to no degeneracy in the ground state when using the deterministic random model, there is no need to find the results for 3 repeats of the single copies (same resources as 3 connected copies) as they will be similar. The total height of the bars on the right show the fraction correct for three copies connected with links $J_F^{(p)}=J_F^{(\text{min})}$.  This is broken down into the three possible cases: all three copies correct (dark blue, lowest), two of three copies correct (green, middle) and one of three copies correct (yellow, top).  The error bars (black) at the top of each bar show that the effect is much larger than statistical effects.
These results clearly show how the frustration tends to break one or more copies: the bar for all three connected copies correct is always lower than the single copy fraction correct.  It also shows that the frustration tends to keep at least one copy correct, even when a single copy is broken by the reduced precision: the total fraction correct for three connected copies is always higher than the single copy fraction correct.

\begin{figure}[!tb]
    \centering
    \includegraphics[width=0.8\columnwidth]{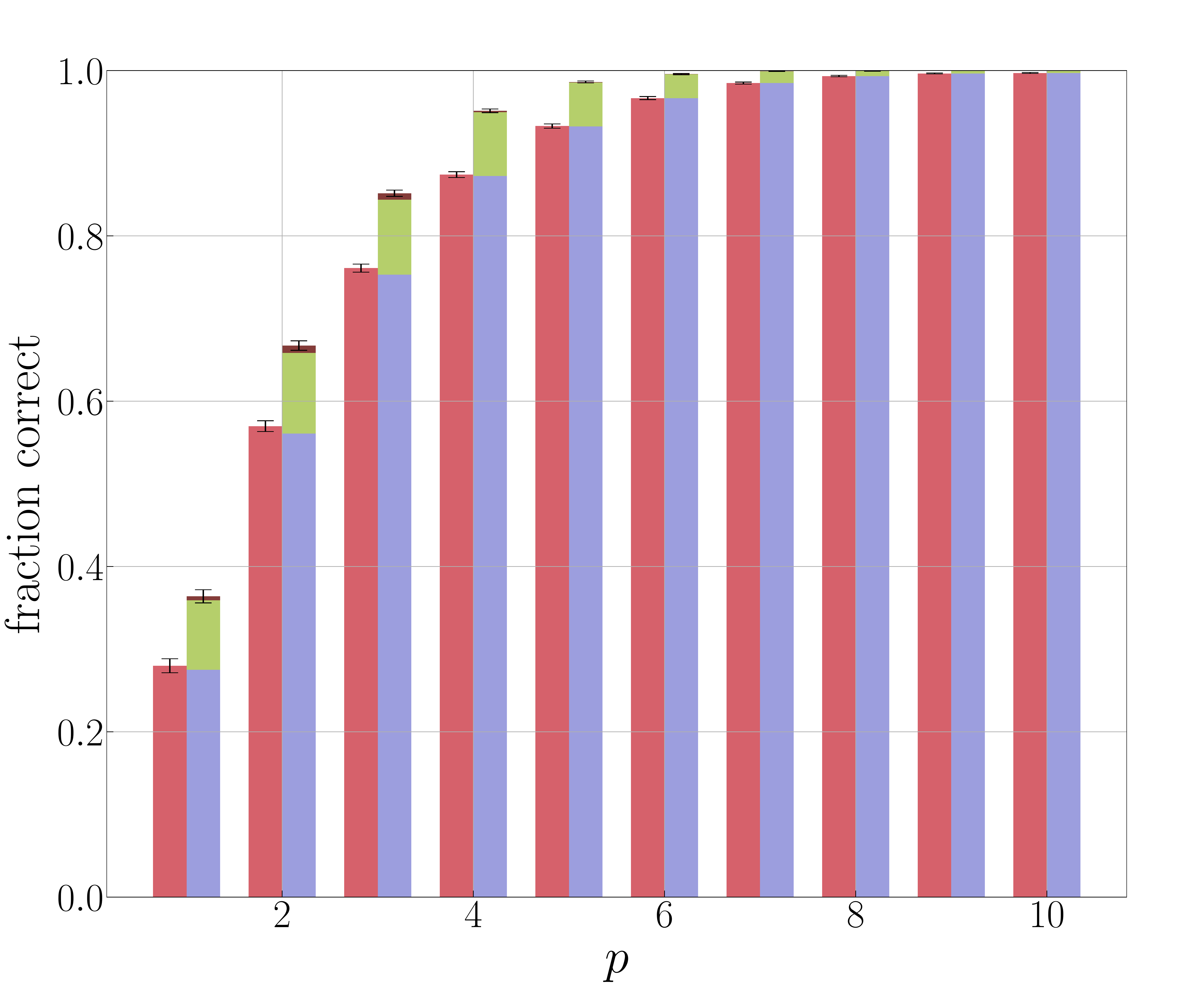}
    \caption{Fraction correct versus precision $p$ for $10^4$ instances of $5$-qubit spin glasses. Left bars  show fraction correct for a single copy (red). Right bars show fraction correct broken down into cases where: both single copy and three copies are correct (bottom, light blue); the single copy is incorrect but the three copies have at least one correct (middle, light green); the three connected copies are incorrect but the single copy is correct (top, dark red).}
    \label{fig:sk_if_hist}
\end{figure}

Since the improvement is an average effect, it is also interesting to ask how often connecting three copies anti-ferromagnetically breaks the ground state for a given precision $p$ when it is correct for a single copy. Figure \ref{fig:sk_if_hist} shows this break down of the three copy results.
The left hand bars show the 1 repeat single copy fraction correct (light red) as in figure \ref{fig:frac_corr_hist}. Note that the 3 repeat and 4 repeat single copy fraction correct was also calculated, but due to the low/no degeneracy of the ground states, did not show any significant difference.

The right bars now show: the cases where the single copy is correct and the three copies have at least one correct (bottom, light blue); the cases where the single copy is incorrect but the three copies have at least one correct copy (middle, light green); and the cases where the single copy is correct but the three copies are all incorrect (top, dark red).
The sum of the light blue and light green bars is the same as the total height of the right hand bars in figure \ref{fig:frac_corr_hist}, i.e., all the cases where the three copies give the correct result.  The sum of the dark red and light blue bars is the same as the height of the light red left hand bars, i.e., all the cases where the single copy is correct.

Though the increase in fraction correct is small, running a single (disconnected) copy as well as the 3 connected copies (on a machine one third the size and therefore cheaper) and comparing the candidate solutions to choose the best, allows us to include the dark red section of the right hand bars to our total fraction correct. 
\begin{figure}
    \centering
    \includegraphics[width=0.8\columnwidth]{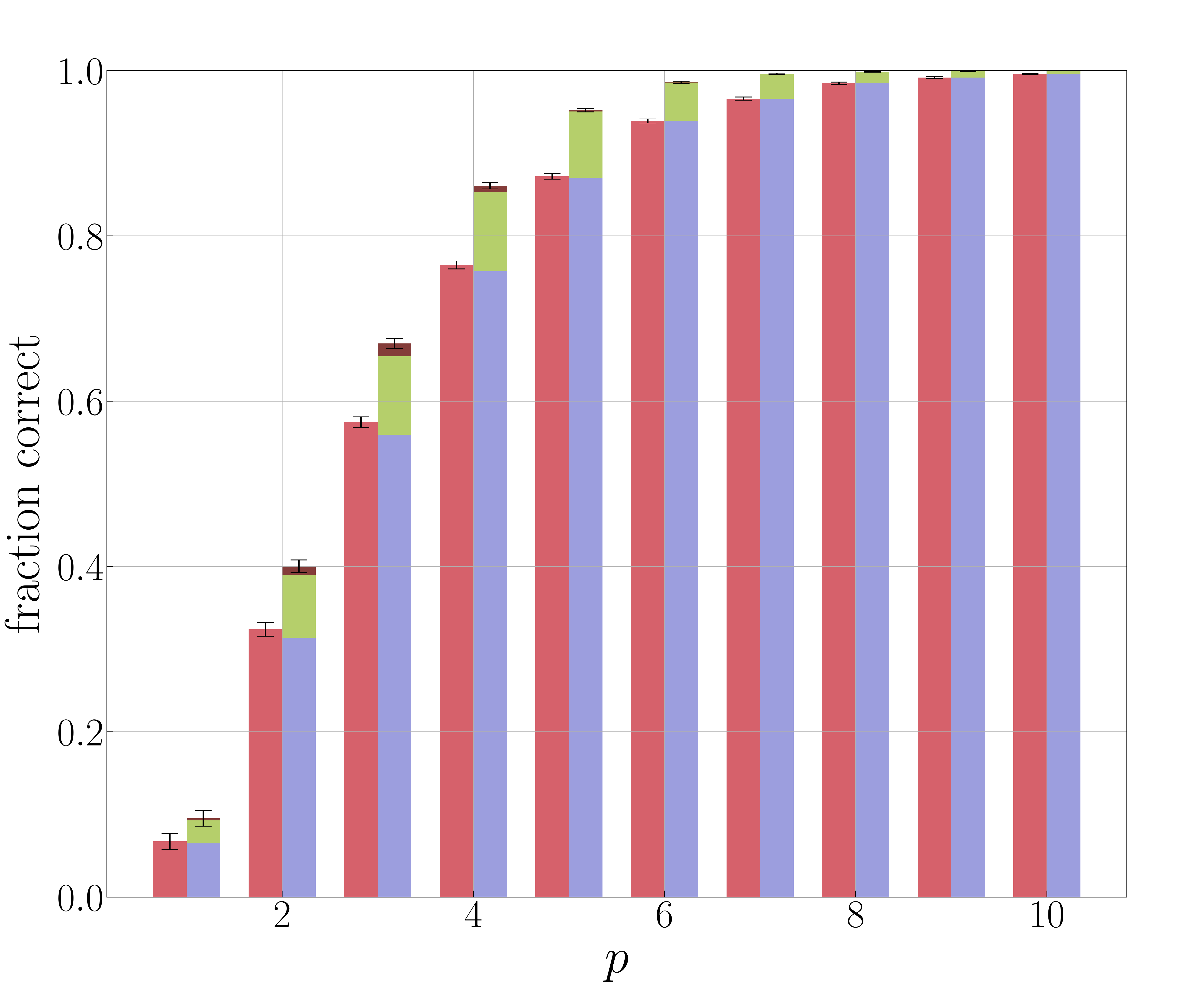}
    \caption{ As figure \ref{fig:sk_if_hist} for $10^4$ instances of $9$-qubit Ising spin glasses.}

    \label{fig:9_sk_if_hist}
\end{figure}
The equivalent figure for $9$-qubit SK spin glasses is shown in figure \ref{fig:9_sk_if_hist}.  It is similar to the $5$-qubit results in figure \ref{fig:sk_if_hist}, but shows how larger system size problems are broken more easily by low precision, as would be expected.

\subsection{The random error model}

For the deterministic error model, once the error has been applied, the ground state we find is either correct or incorrect. As the error is fixed, repeats do not help to find the correct ground state.
To simulate the random error model, we take 10 error samples for each Ising model instance. In general, the single copy will only be incorrect for some of these errors samples, so the fraction correct we calculate is an average over the samples and the instances. Equivalently, we can then look at the fraction correct as the probability of the single copy being correct or incorrect. This means that as long as there is some probability that the ground state is correct, more repeats can help to find the correct ground state.
If we assume completely uncorrelated errors, we can  then find the fraction correct for three unconnected copies $P(3)$ (the same as three repeats of a single copy) using,
\begin{equation}
    P(3) = 1 - (1-P(1))^3,
    \label{eq:3_copies_prob}
\end{equation} 
where $P(1)$ is the fraction correct for the single copy.

\begin{figure}[!tb]
    \centering
    \includegraphics[width=0.8\columnwidth]{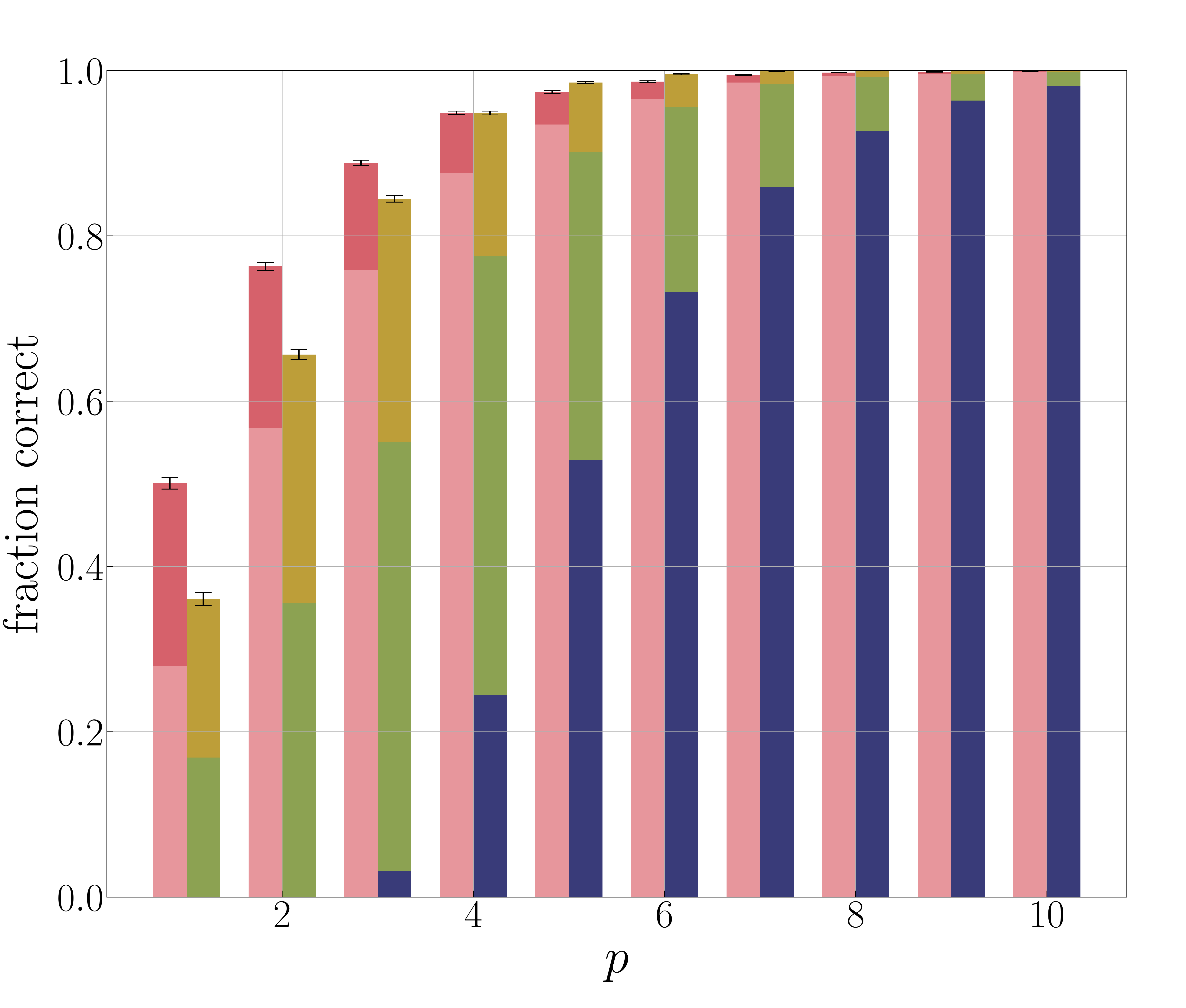}
    \caption{As figure \ref{fig:frac_corr_hist} but for the random error model. The left hand bars are split into the fraction correct of a single copy (light red, lower) and the fraction correct of 3 unconnected copies (red, upper), calculated using equation \ref{eq:3_copies_prob}. 
    }
    \label{fig:frac_corr_hist_rand}
\end{figure}

In figure \ref{fig:frac_corr_hist_rand} the comparison between three unconnected copies (the same as three repeats of a single copy) and three copies connected anti-ferromagnetically is shown in the same format as in figure \ref{fig:frac_corr_hist}.  The fraction correct is now larger for the three unconnected copies, but the three connected copies still show additional improvement for higher precisions $p\gtrsim 5$.
These results show that, for the random error model, the anti-ferromagnetic links provide an additional effect beyond independent repetition, albeit less than in the deterministic error situation.

As we assume completely uncorrelated errors here, we can use equation \ref{eq:3_copies_prob} to calculate the fraction correct for three unconnected copies. However, in hardware it is likely that there would be a mixture of both uncorrelated and correlated errors, meaning the improvement from repeats would be smaller than that seen in figure \ref{fig:frac_corr_hist_rand}.
The best balance between repeats of single copies and connecting multiple copies thus depends on the specific hardware implementation.
%though the extent of this would have to be determined experimentally.
%We expect the effectiveness of repeating single copy runs to reduce for larger sizes, because there is more chance that while one random error is removed, another is introduced, but we have not tested this due to computational resource limitations.
%
\subsection{Spin chains}
% spin chain note:
%
We have also analyzed equivalent data for $5$-qubit spin chains for all the cases in this section. 
We find that the effects are very similar to the spin glasses for all cases, and so do not present the details here.
The similarity in all aspects provides confidence in the generality of our results for other problem Hamiltonians.

%---------------------------------------------------%
\section{Precision improvements}
\label{sec:precision}

We now turn to a quantitative estimate of the improvement in precision, using the deterministic error model applied to the spin glasses.
\begin{figure}[!tb]
    \centering
    \includegraphics[width=\columnwidth]{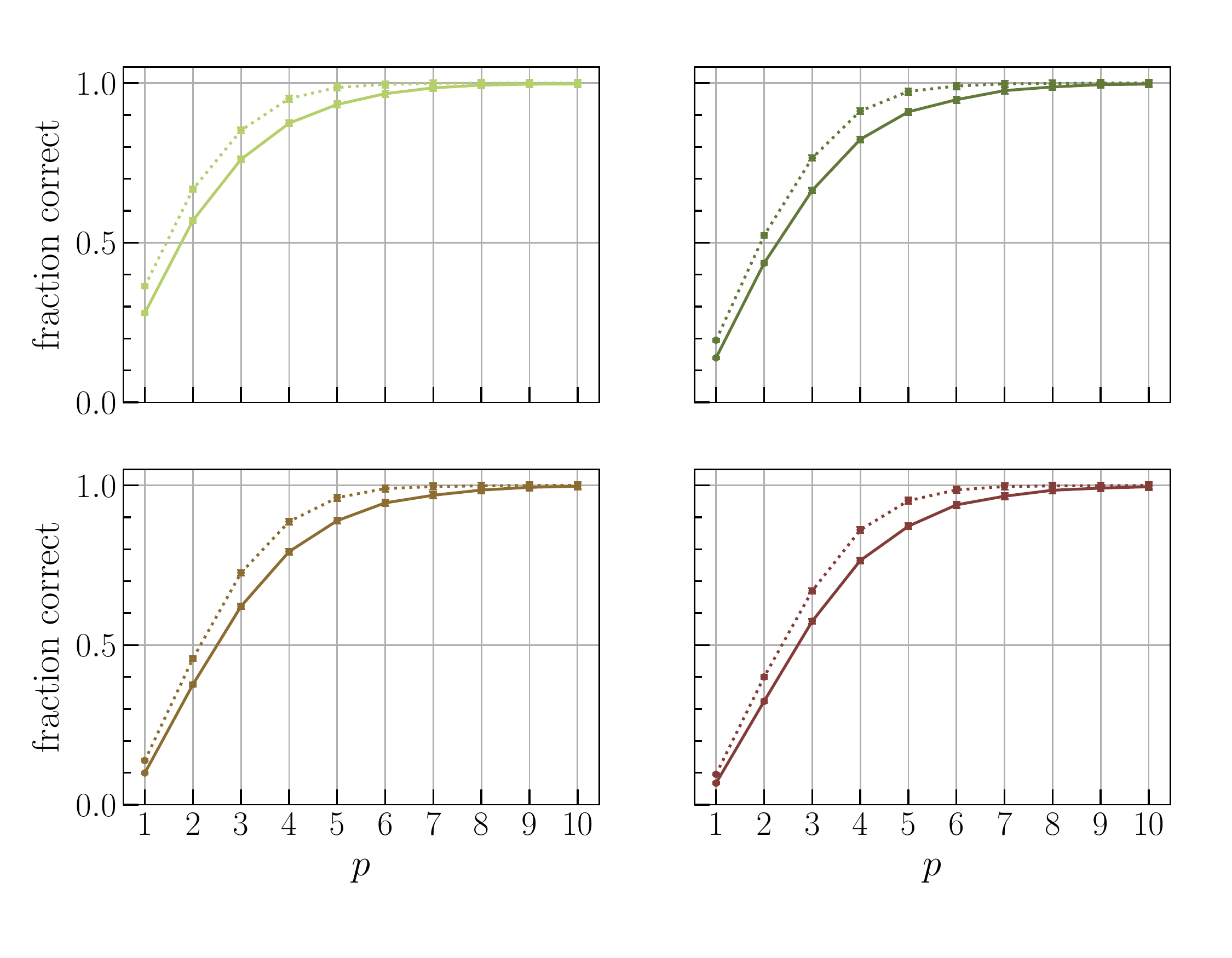}
    \caption{Fraction correct vs precision $p$ for $10^4$ instances of 5 (light green), 7 (dark green),  8 (brown) and 9 (red) qubit spin glasses with four repeats of a single copy (solid) and three connected copies plus 1 separate single copy (dotted).}
    \label{fig:spin_glass}
\end{figure}
Figure \ref{fig:spin_glass} shows the total fraction correct (peak of the bars in figure \ref{fig:9_sk_if_hist}) for single copies (solid lines) and three connected copies plus a separate single copy (dotted lines) for $10^4$ instances of SK spin glasses for $n=5$, $6$, $8$ and $9$.  Not shown, $n=7$ is very similar.  The limit we can compute using reasonable computational resources is $n=9$, for which three copies requires 27 qubits.

The single copy fraction correct approaches unity around $p=n+1$. 
As noted previously, the intuition for this is that there are $2^n$ different possible states for $n$-qubits, i.e., $n$ qubits cannot represent higher precision outcomes than $p=n$.
For the three connected copies (plus a separate single copy), the fraction correct (dotted lines) tends to approach unity sooner.  In other words, the three connected copies still find the correct solution at lower precision than for a single copy.  This is the effect we are looking for, to protect against lack of precision in the hardware.

%--------------------------------------------%
\subsection{Quantifying improvements}

\begin{figure}[!tb]
    \centering
    \includegraphics[width=\columnwidth]{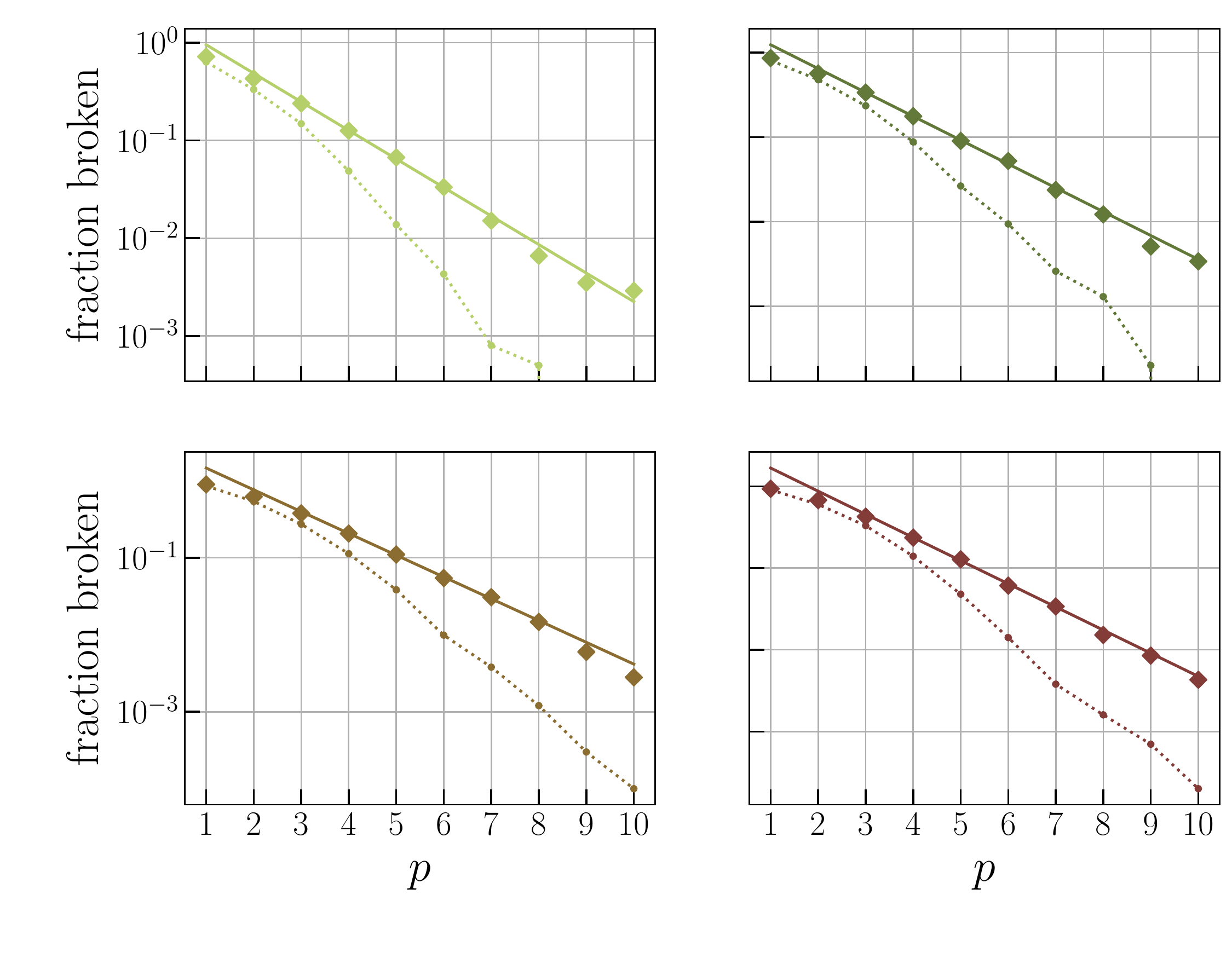}
    \caption{Same data as figure \ref{fig:spin_glass}, but  plotted as fraction broken $= 1 -$ fraction correct on a log scale,
    with a linear fit to the single copy data (solid lines) excluding the first two points ($p=1,2$) from the fit.}
    \label{fig:frac_fits}
\end{figure}

\begin{figure}[!tb]
   \centering
    \includegraphics[width=0.8\columnwidth]{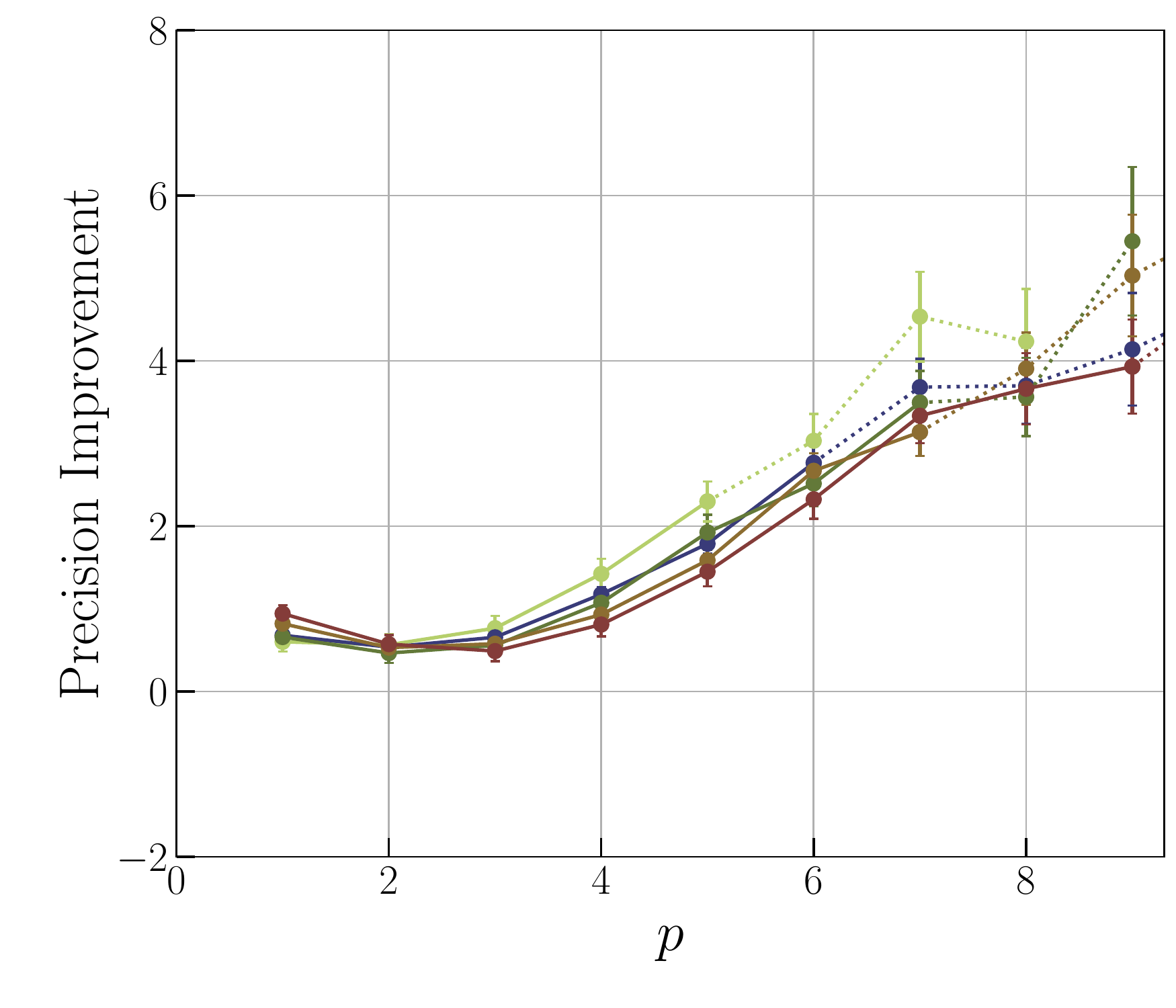}
    \caption{Precision improvement in bits versus precision, for $10^4$ instances of 5 (light green) 6 (blue), 7 (dark green), 8 (brown) and 9 (red) qubit spin glasses. Precision improvement was calculated between a linear fit to the 4 repeats of a single data and  the 3 connected copies plus 1 single copy data at the same value fraction correct. Data below $p=n$ is plotted as solid lines. Data above $p=n$ is plotted as dotted lines. 
    %%%
    %%%
    }
    \label{fig:improves_glass}
\end{figure}

To quantify the improvements we obtain by connecting three copies anti-ferromagnetically plus a single copy, we calculated the difference in precision $p$ between the single copies and connected copies plus single copies at the same fraction correct. We called this the precision improvement.  
In figure \ref{fig:spin_glass}, this is the horizontal distance between the dotted and solid lines.
However, as the data points on the dotted line (three-connected-copies-plus-single-copy) do not in general have corresponding data points on the solid line (single-copy) at the same fraction correct, we used the data points for three-connected-copies-plus-single-copy data and extrapolated the single-copy data.  The extrapolation uses an exponential fit of the form $f(p) = A \exp(-bp)$ where fraction correct $=(1-f(p))$, in order to estimate an effective precision for the single copy, at the corresponding value of fraction correct. All fits exclude the single copy data points at $p=1$ and $p=2$, which exhibit non-exponential effects due to the low precision. These fits are plotted in figure \ref{fig:frac_fits}, alongside the single-copy and three-connected-copies-plus-single-copy data on a log fraction broken $= (1-$ fraction correct$)$ vs precision plot.
The difference between the value of the precision at the same fraction correct (horizontal displacement) is the \emph{precision improvement}, plotted in figure \ref{fig:improves_glass}. The error in the variables $A$ and $b$ (arising from the fit) and the error in fraction correct (from the three-connected-copies-plus-single-copy data points), were combined via a functional approach \cite{Hughes} in order to calculate the error in the single-copy precision.  This is the error shown by the error bars in figure \ref{fig:improves_glass}.

It can be seen in figure \ref{fig:improves_glass}, for each size $n=5$ to $9$, between precision $p=2$ to around $p=7$, as precision increases, the precision improvement also increases, from less than 1 at $p=2$ to around 3 at around $p=7$. 
At $p=1$ the precision improvement appears to be larger than at $p=2$, but this is due to the exponential fit not matching the data at this very low precision. 
Data points for $p\gtrsim 6$ have larger error bars due to the small number of incorrect single copy counts at high precision.

The levelling off of the improvement suggested by the data for higher $p$ is expected for the finite resources of three copies.  
As this work is a numerical study on small problem sizes, we cannot reliably extract trends for larger problem sizes. However, the results presented here offer proof-of-concept that the method can provide several extra bits of precision by using three linked copies compared with a single copy.

%----------------------------------------------------%
\section{Continuous-time dynamics}\label{sec:QWs}

The results in the previous sections are obtained by considering the problem Hamiltonian only.  This provides good estimates for the outcome using adiabatic processes that remain in the ground state throughout.  For more dynamic processes that populate excited states, it is also necessary to check that actually running a quantum anneal or quantum walk does not introduce adverse effects that negate the precision improvement.
\begin{figure}[!tb]
   \centering
    \includegraphics[width=0.5\textwidth]{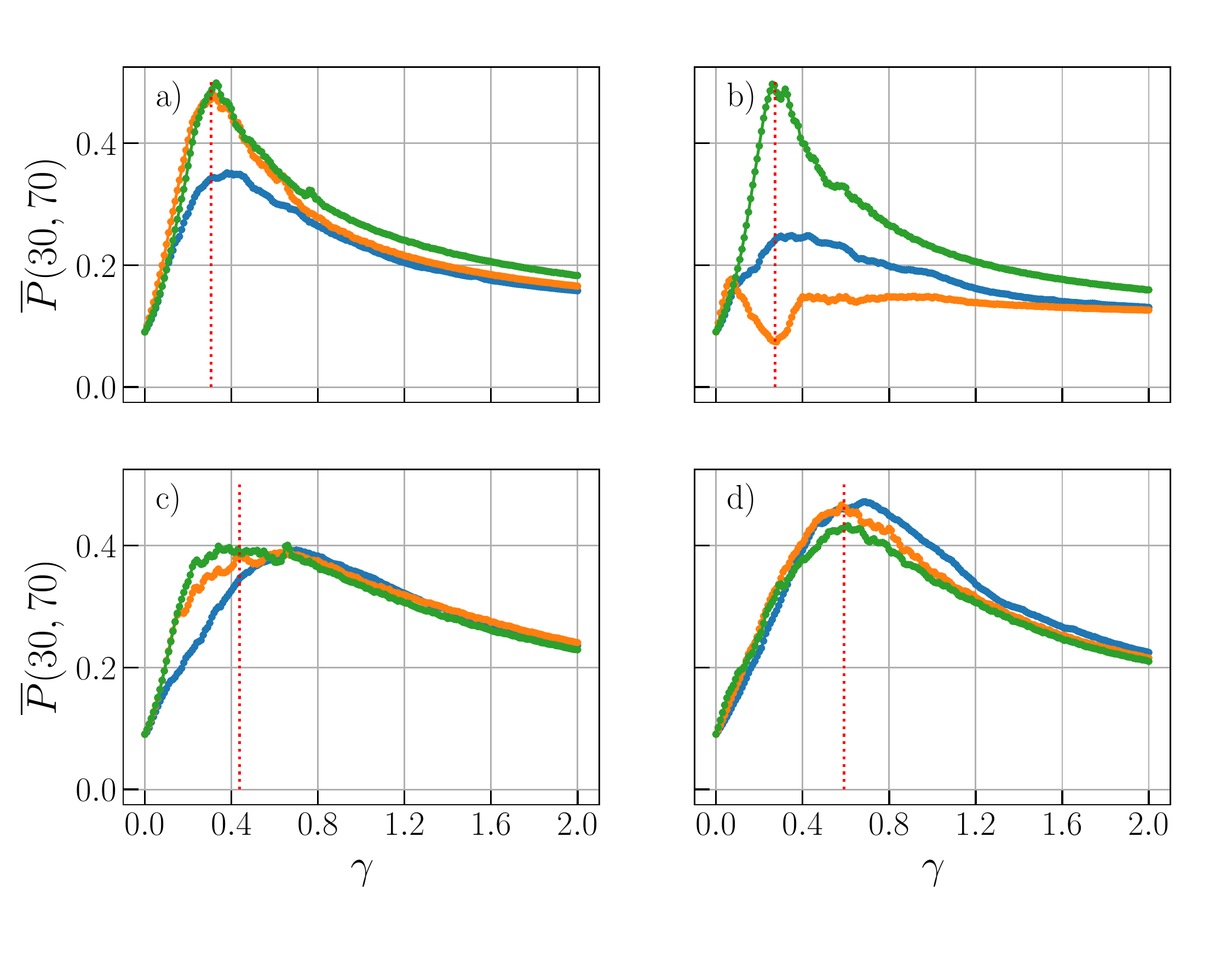}
    \caption{Average success probability equation \eqref{eq:Time_av}, $P(t,\Delta t)$ for $t=30$, $\Delta t= 70$ versus $\gamma$, for four instances a) to d) of a  $5$-qubit spin glass for: no links ($J_F=0$) and no precision reduction (green); with $J_F=0$ and precision set to $p=3$ (orange); with $J_F\simeq J^{\text{min}}_F$
    and $p=3$ (blue). The precision reduction used the deterministic random error model.  
    Each instance a) to d) has an incorrect ground state for a single copy but is correct for at least one of three connected copies for $p=3$.
    The optimal value of $\gamma$ for the case with no links and no precision reduction, (peak of the green line) is shown as a vertical red dotted line.
    For reproducibility, the unique I.D.'s of the instances a) to d) can be found in section \ref{sec:numeth}.}
    \label{fig:QW_ex}
\end{figure}
%
%For adiabatic quantum computing, as the system will always remain in the ground state, we don't expect the connected copies to affect the performance.
%
%For imperfect adiabatic techniques such as quantum annealing, or for quantum walks, excited states other than the ground state are populated.  
We tested our methods with quantum walks, as the time-independent Hamiltonian is easier to simulate, and it also tends to populate more excited states than an approximately adiabatic dynamics.

For quantum walks, the key parameter is the hopping rate $\gamma$ that determines the relative strengths of the driving and problem Hamiltonians.  For practical applications, the success probability must not be very sensitive to the exact value of $\gamma$, because there is in general no efficient way to determine the optimal value without solving the problem itself.  For the spin glasses, \cite{Callison2019} discusses in detail how to estimate a suitable heuristic value of $\gamma$.
It turns out that the spin glasses have a fairly broad range of suitable $\gamma$ values, as illustrated by the green curves in figure \ref{fig:QW_ex}.  The optimal value determined in \cite{Callison2019} is indicated by the vertical red dotted lines.  

There are a few subtleties to correctly comparing the three-copy performance with the exact single-copy performance in a quantum walk setting.  The average success probability is defined as the overlap of the quantum state with the solution state, time averaged as per equation \eqref{eq:Time_av}.  For three copies, there are many different states that give one or more correct copies, and not all of them are ground states.  We define the success probability to be the sum of the probabilities of obtaining any of these states -- since what we are interested in is correct solutions, not ground states -- and then we time averaged in the same way.  For the exact single copy case, the probability of obtaining a correct solution from three runs can be calculated as follows, where $P(y)$ is the probability of succeeding at least once in $y$ runs.  We have $P(3) = 1-(1-P(1))^3$, i.e., subtract the probability of failing all three runs from unity.  This is the correct quantity to compare with three copies, as it uses the same number of qubits in total, and can be run in the same time (by just setting the anti-ferromagnetic links to zero).

As well as the exact three-run success probabilities plotted in green in figure \ref{fig:QW_ex}, also shown are the success probabilities for the same spin glass instances at reduced precision $p=3$ both with (blue) and without (orange) the anti-ferromagnetic links close to the minimum precision
(using the deterministic random error model).  
The orange lines thus show the difference the lack of precision makes to single copies compared to the green exact success probabilities.  The blue lines show how anti-ferromagnetic links change this for the same lack of precision.
Each of the four instances were chosen so that, at $p=3$, their ground state was incorrect when only a single copy was used, but they had at least one copy with a correct ground state, when three connected copies were used.

The four instances plotted in figure \ref{fig:QW_ex} illustrate the range of behaviours we observe.  
In these examples, the probability of obtaining the correct solution state is sometimes higher, sometimes lower and sometimes similar, but in all cases, the broad peak remains, and the optimal value of $\gamma$ would provide a good performance, except for one single copy instance, figure \ref{fig:QW_ex}(b).
This shows that the dynamics and heuristic parameter estimation are not significantly impacted by the anti-ferromagnetic links joining multiple copies.
Moreover, given that the success probability is in any case not sensitive to the exact value of $\gamma$, it is not necessary to use any error mitigation techniques to increase the precision for the driver Hamiltonian settings, for solving this type of problem. 

Characterizing the effects of limited precision and anti-ferromagnetically linked copies on the outcomes of quantum walk computation requires significant work to unravel the contributions from excited states that also provide correct solutions.  A full analysis will be left for future work.

%----------------------------------%
\section{Conclusion and open questions}\label{sec:conc}

In this work we have introduced, and provided a proof-of-concept numerical analysis of, a set of improvements to the error suppression part of the error correction scheme first introduced in \cite{Young2013}.
Recognising that only one copy needs to be correct in a quantum optimization setting allows us to use anti-ferromagnetic links instead of ferromagnetic.
The advantage of anti-ferromagnetic links is that they penalise configurations in which all copies have the same error, improving the chance of at least one copy remaining correct, in a repetition-code multiple-copy model.  
Furthermore, only weak anti-ferromagnetic links are needed to achieve this error suppression, making our method fully scaleable.

We also confirmed that introducing linked copies does not significantly change the parameters and performance of quantum walk dynamics used to solve the spin glass instances. %
In other words, heuristics for setting parameters based on single copies can be used without modification for the multiple linked copies. Interpolation between quantum walks and AQC \cite{Morley2017} suggests this may hold for quantum annealing parameters, too, although we have not tested this.  

Applying these methods to a setting in which only limited precision is allowed for the fields and coupling strengths, we showed a proof-of-concept that anti-ferromagnetic links can compensate for the errors introduced by lack of precision, effectively gaining several bits of precision for three connected copies.  If this extends to practical sizes, it provides a way to scale quantum annealing systems beyond the barrier of fixed precision in the controls for the fields and couplings between the qubits.  Our work highlights the importance of incorporating frustration into quantum annealing architectures.  This took the form of anti-ferromagnetic triangles in our examples here; we have also tested larger odd and even loops to confirm that frustration is the key element.

Further work is needed to extend these results analytically, to larger sizes, and potentially to concatenating the repetition codes in analogy to gate model error correction.  The mean field methods in \cite{Matsuura2016,Matsuura2017a} can potentially be adapted to this setting to provide indications of the likely scaling.  Preliminary study supports our numerical finding that the quantum walk hopping rate does not change, an important factor for scalability.  Mean field methods may also allow the interpretations in \cite{Vinci2018} in terms of effective temperature reduction, or energy scale enhancement, to be applied to our method, facilitating a direct comparison of the methods and their suitability for different types of errors.

Practical demonstrations of the precision improvement could be done on hardware with a suitable native graph -- results for longer spin chains indicate that minor embedding can reduce rather than increase precision, which could lead to inconclusive results. 
However, the experimental results for nested QAC in \cite{Vinci2015,Vinci2016a} suggest it should be possible to extract some indications of the practical validity of the method using current DWave systems.

Furthermore, it is an open question whether the techniques proposed here could be applied to gate-model machines through QAOA (quantum approximate optimisation algorithm), given the recently highlighted connections between QAOA and quantum annealing \cite{Brady21a}.  
Whether these techniques can be extended beyond classical optimization problems and classical repetition codes to fully universal quantum error correcting codes, e.g., for quantum simulators, where quantum Hamiltonians are evolved continuously in time, is also an interesting open question. 
The addition of weak anti-ferromagnetic links effectively rewards finding good quality solutions which differ by large Hamming distances. This may have application beyond error mitigation in hybrid algorithms, or for problems where a diverse set of solutions are desired.

\acknowledgements
NC and VK were supported by EP/L022303/1 and impact acceleration funding associated with this grant.
NC was supported by EPSRC fellowship EP/S00114X/1.
JB was supported by a UK EPSRC funded DTG studentship, project reference 2214392.
AC was supported by the EPSRC UK Quantum Technology Hub in Computing and Simulation (EP/T001062/1).

%%%% Refs after appendices in RevTex...%%%%%
\bibliography{AQCQEC.bib}

\end{document}